\documentclass[sigconf]{acmart}

\copyrightyear{2019}
\acmYear{2019} 
\setcopyright{iw3c2w3}
\acmConference[WWW '19]{Proceedings of the 2019 World Wide Web Conference}{May 13--17, 2019}{San Francisco, CA, USA}
\acmBooktitle{Proceedings of the 2019 World Wide Web Conference (WWW '19), May 13--17, 2019, San Francisco, CA, USA}
\acmPrice{}
\acmDOI{10.1145/3308558.3313432}
\acmISBN{978-1-4503-6674-8/19/05}
\usepackage{subcaption}

\usepackage{booktabs} 

\newcommand{\xhdr}[1]{\vspace{1mm}\noindent{{\bf #1.}}}

\usepackage{xcolor}

\begin{document}
\title{Goal-setting And Achievement In Activity Tracking Apps: A Case Study Of MyFitnessPal}

\author{Mitchell L. Gordon}
\affiliation{\institution{Stanford University}}
\email{mgord@cs.stanford.edu}

\author{Tim Althoff}
\affiliation{\institution{University of Washington}}
\email{althoff@cs.washington.edu}

\author{Jure Leskovec}
\affiliation{\institution{Stanford University}}
\email{jure@cs.stanford.edu}

\begin{abstract}
Activity tracking apps often make use of goals as one of their core motivational tools. There are two critical components to this tool: \textit{setting} a goal, and subsequently \textit{achieving} that goal. Despite its crucial role in how a number of prominent self-tracking apps function, there has been relatively little investigation of the goal-setting and achievement aspects of self-tracking apps.

Here we explore this issue, investigating a particular goal setting and achievement process that is extensive, recorded, and crucial for both the app and its users' success: weight loss goals in MyFitnessPal. We present a large-scale study of 1.4 million users and weight loss goals, allowing for an unprecedented detailed view of how people set and achieve their goals. We find that, even for difficult long-term goals, behavior within the first 7 days predicts those who ultimately achieve their goals, that is, those who lose at least as much weight as they set out to, and those who do not. For instance, high amounts of early weight loss, which some researchers have classified as unsustainable, leads to higher goal achievement rates. We also show that early food intake, self-monitoring motivation, and attitude towards the goal are important factors. We then show that we can use our findings to predict goal achievement with an accuracy of 79\% ROC AUC just 7 days after a goal is set. Finally, we discuss how our findings could inform steps to improve goal achievement in self-tracking apps.

\end{abstract}

\maketitle

\section{Introduction}
The purpose of an activity tracking app is to help users better understand their behavior. Health-focused apps and fitness devices are the most prevalent, often tracking activities like exercise, eating, and heart rate. Beyond physical health, other types of behaviors are becoming increasingly popular to track as well; for instance, iPhones now come pre-installed with an app that tracks screen time spent in other apps.

In many cases, users wish not only to observe their behavior, but also to improve it \cite{rooksby2014personal, gowin2015health}. Activity tracking apps often aim to help their users do this, and the guidance they provide can take many forms, ranging from general advice and tips from experts, peer pressure from social networking features, reminders or notifications that ask the user to take a specific timely action, and enforcement of explicitly articulated goals \cite{schoeppe2016efficacy}. We think of all these mechanisms as methods to promote behavioral change \cite{bandura1977self}. They play an important role in people's well-being, akin to the many methods for behavior change in the offline world that one is familiar with, which might include requests or demands from loved ones, medical professionals, bosses, or in the most extreme cases, governmental intervention via fines and criminal punishment.

A nearly universal feature in activity tracking apps, goals are one of the most popular and effective methods for behavioral change \cite{mercer2016behavior, shilts2004goal}. They take many forms. For instance, Apple Watch has three ``rings'' for the user to fill in, each representing a different daily behavioral goal: one for standing 12 times, one for total calories burned (set by the user), and one for 30 minutes of exercise. Other apps have focused on moderating digital behaviors by setting goals for the amount of time spent on social networking sites \cite{althoff2017online}. Perhaps the most well-known goal is walking 10,000 steps per day assigned by many fitness-focused activity trackers \cite{choi2007daily}.

Goals in activity tracking apps involve both the setting of a goal (the selection of a desired outcome) and also the achievement of that goal (the process of working towards the selected outcome). Setting a good goal is critically important for users' success: too hard, and the user will become frustrated and give up. Too easy, and they won't achieve as much health benefit as they could have \cite{strecher1995goal}. There has been interesting recent work on goals in activity tracking apps \cite{wen2017evaluating,mansi2015investigating,olander2013most,bodenheimer2009goal,donaldson2009using,pearson2012goal,laing2014effectiveness}, but it remains a topic where there is still much to be understood -- particularly on the issue of setting a good goal, since it can be difficult to find records of the process by which goals are both set and subsequently achieved, along with the behaviors that led to those goals being achieved. This has led to the problem that, despite the importance of goal-setting and achievement in activity tracking apps, users often receive limited guidance on selecting a goal that they will actually meet. While most activity trackers use general guidelines from relevant domain expertise combined with demographic information \cite{davis2016health} to help users make a choice, after the initial goal is set they often provide limited on-going advice as to whether that goal choice is turning out to have been a good one. Some work has gone further and used current trajectory towards a goal to intervene. For instance, they can notify a user if they are not on track to meet their goal by a certain time frame if they keep that trajectory, or are consistently not meeting daily goals \cite{normand2008increasing,herrmanny2016supporting,dishman2009dose,op2013opportunities, adams2013adaptive}. However, these methods do not consider a holistic view of the many important behaviors that lead to a goal being met, and make simplistic assumptions about linear progressions. 

\xhdr{Present Work: Weight Loss Goals in MyFitnessPal}
In this paper we consider the goal-setting aspects of activity tracking apps by focusing on a setting where detailed information on both how goals are selected and many related behaviors while attempting to achieve those goals can be studied. Our setting is the process of selecting and working towards weight loss goals in MyFitnessPal. This process has a clearly defined structure: first, upon creating a MyFitnessPal account, all users select weight loss goals, and do so without any guidance from the app. Users optionally provide a free-text reason for their selection of that goal. Users then may then use the app to log their food intake and weight to help them track their progress towards that goal. Meeting their weight loss goal is important for their health: achieving a self-determined weight goal has been linked to long-term weight loss, while the failure to reach a self-determined weight may discourage someone's belief in their ability to control their weight, resulting in abandonment of weight loss behaviors and weight re-gain \cite{elfhag2005succeeds, linde2004unrealistic}.

MyFitnessPal weight loss goals thus have the key ingredients we need: they take place in a goal-setting and achievement process carried out by users of a activity tracking app who are working towards meeting that goal, users use the app to help guide their lower-level behaviors that will help them to meet that goal, and given the nature of activity tracking, all of this is logged, making the analysis possible. It also serves as an instance of a broader type of goal-setting, familiar from the off-line world as well as the on-line world, in which people aim to lose weight by setting a weight loss goal. 

Both the process of goal-setting and the process of losing weight has been studied extensively in medical literature \cite{elfhag2005succeeds}. By contrast, because of our interest in the issue of goal-setting and achievement in activity tracking apps, we study the process from the perspective of early detection of goals that will never be met: we ask which goals users set, how users behave shortly after they set them, and how those behaviors lead to those goals being achieved. Because we are interested in exploring how activity tracking apps may be able to help users select better goals, we focus specifically on behavior that takes place during the first 7 days after a goal is set, and examine whether we might be able to detect early-on that a long-term weight loss goal will never be met. Such early-detection would enable an intervention warning a user that they are working towards a goal they are unlikely to meet, which means we may be able to encourage them to re-think a more realistic goal and prevent them from spending too much time working towards and becoming invested in that goal, only to ultimately become discouraged and re-gain weight because they did not meet it. This could potentially be considered a ``just-in-time intervention'' \cite{nahum2017just}.

\xhdr{Present work: Main Results} We analyze 2.8 million weight loss goals with 44.6 million weights logged over a period of three years. We begin by validating our dataset through a comparison of the goals the people set in MyFitnessPal to prior studies from before the age of activity trackers, finding that women set more ambitious goals than men and younger adults set more ambitious goals than older adults (Section 4). We also use the unprecedented size of our dataset to show that some findings in prior work are likely due to selection effects. We then analyze goal completion rates and duration, finding that men and older adults are more likely to meet their goals (even when accounting for goal difficulty), but that overall completion rates are low (Section 5).

We then turn to the problem of early detection of goals that will never be met (Section 6). We investigate user's behavior over the first seven days after they set their goal. We find that, contrary to what some prior clinical studies might suggest, there's no such thing as ``too much'' early weight loss when it comes to meeting your goal (Section 6.1), and that people who log their weight more frequently within the first week are more likely to meet their goals (Section 6.2). We show that more calories reported per day during the first 7 days leads to a lower likelihood of meeting goals, but only for users who are committed loggers; surprisingly, for many users, logging more calories actually indicates a higher propensity to meet a weight loss goal (Section 6.3). We then turn to the motivations behind the goals that people set, introducing a novel application of topic modeling algorithms to identify four primary motivators and show how achievement rates vary with these motivators (Section 6.4).

Finally, we show that whether a user will achieve their target weight can be predicted just based on the initial behavior during the first few days after the target is set. We build a machine learning model to predict target achievement with promising accuracy. We conclude with a discussion of how these results can be translated into actionable suggestions for activity tracking apps as well as traditional offline weight loss programs.

\section{Related Work}

\xhdr{Goal-setting Theory and Weight Loss}
There is a vast body of work discussing the mechanics and psychological aspects of goals, as well as how goals can be used for health-related behaviors. Locke and Latham's seminal work summarizes empirical research on goal-setting theory \cite{linde2004unrealistic}. Other researchers have focused on how those findings apply specifically to health goals \cite{gollwitzer1998emergence,strecher1995goal}.

As with goal-setting theory, there is also a vast body of work discussing the mechanics of weight loss and weight loss goals. As weight loss goals tend to focus on longer term weight loss (as discussed later, we find most people want to lose significant percent of their body weight), literature relating to longer-term weight loss and maintenance, such as Elfhag and Rossner conceptual review of the subject, is most relevant to our work \cite{elfhag2005succeeds}. Other work has focused on characterizing weight loss goals, both in terms of what goals people set and which goals are associated with the best outcomes \cite{jeffrey1998smaller,linde2004unrealistic,williamson1992weight,foster1997reasonable,linde2005weight}. 

Our work first verifies that findings from these smaller-scale offline studies generally apply to our activity tracking dataset in terms of which goals people choose to set and their likelihood of meeting them, while also using the unprecedented scale of our dataset to present new insights. We then extend existing work by specifically investigating early behaviors that predict whether or not a user will meet their weight loss goal in an activity tracking app.

\xhdr{Implementing Goals in Activity Trackers}
Activity trackers provide users with an environment to both set goals and track their progress towards their goals. In the best case, they might also be described as systems to help people implement their goals \cite{gollwitzer1999implementation}. Some work has looked at different higher-level strategies for implementing goals in self-trackers, typically through qualitative small-scale studies \cite{dennison2013opportunities,munson2012exploring,consolvo2009goal}. Other work has investigated the role that goals play in changing behavior in self-trackers \cite{epstein2016beyond, epstein2016beyond,cordeiro2015barriers,nothwehr2006goal}. Rather than exploring the best way to implement goals in activity trackers, our work is a case study of one of the world's most popular implementations: MyFitnessPal.

\xhdr{Adaptive Goals}
While goals in self-trackers frequently take into account one's current size and weight, personalized or adaptive goals that take behavior in to account are used sparsely in practice 
\cite{op2014tailoring}. 
Some researchers have sought to assign adaptive fitness goals based on a particular user's historical data, in order to improve a goal that is too difficult or too easy. These typically employ simple algorithms that raise goals that the user is easily achieving, and lower goals that the user is struggling to achieve 
\cite{normand2008increasing,herrmanny2016supporting,dishman2009dose,op2013opportunities,adams2013adaptive}. 
MyBehavior, in a study of 17 users, suggested exercise or actions to take, but not goals \cite{rabbi2015mybehavior}. Hermanny \textit{et al.} explored using heart rate variability to set goals \cite{herrmanny2016using}. These adaptive goals can be considered part of a new class of ``just-in-time interventions'' \cite{nahum2017just}. 

Critically, existing work has focused on measuring progress directly related to the goal (e.g. steps for step count goals, calories for calorie goals). Our work builds on a large dataset covering many behaviors related to achieving the goal and findings from work in goal-setting theory to understand whether a goal will be met by taking a holistic view of a user's behavior. We are aware of one other paper which aimed to predict whether a weight loss goal will be met; however their model relied upon users having already been working towards their goal for at least two months and to have logged a large number of weights, meaning it was only applicable to less than 1\% of the total number of users in their dataset \cite{velivckovic2018cross}.

\section{Dataset Description}
This section describes the mechanics of the activity tracking app MyFitnessPal, the dataset used in this paper.

\xhdr{The Mechanics of MyFitnessPal}
MyFitnessPal enables its users to track many health-related behaviors. Users can log their food intake, their exercise, and their weight. Users can also set health and behavioral goals. There are three main goals in the app: total weight loss, weekly weight loss, and calories per day. In this paper, we focus on the total weight loss goal: the weight that the user would like to achieve. All users enter a total weight loss goal upon creating a MyFitnessPal account. The app does not provide any guidance when a user is selecting their total weight loss goal. Users also decide upon a weekly weight loss goal between 0 and 2 lbs per week, in increments of half a pound. The default value for this field is 1 lb per week. Using this weekly weight loss goal and demographic information entered by the user (such as current weight and height), the app then automatically assigns the user a calories per day limit. Each food that a user logs is counted against this limit, while exercise is counted as burning off some of those calories.

To log a food item, users type the name of what they ate into the app's search bar, which will then return a list of matches from which the user can choose. Many of these items are branded and will contain full nutritional content supplied by their manufacturer (e.g. a McDonald's Big Mac or Trader Joes Granola). Other generic items, such as eggs or chicken breasts, will also contain nutritional content. If the user does not find a suitable match in the app's database, they can optionally log their item manually and choose which, if any, nutrient content to provide. There are two methods for logging a weight. First, a user can manually enter their current weight into the app whenever they would like. Second, users can connect MyFitnessPal an many internet-connected ``smart'' scale, which automatically transfer weights they record to the app.

Progress towards the calories per day limit is prominently displayed on the main page of the app. Users can easily view their weight loss progress by clicking on a large icon in the bar at the bottom of the screen. 

\begin{table}[t]
\begin{center}
\begin{tabular}{ l | l }
    \hline
    Characteristics & Value \\
    \hline
    \# of users studied & 1,413,431 \\
    Observation period & August 2014 - April 2017 \\ 
    Weights logged & 44,568,749 \\
    Weight goals set & 2,803,073 \\
    Food items logged & 8,785,560,832\\
    Free-text goal justifications & 56,014 \\
    Mean age & 38.69 \\  
    \% users female & 71.9 \\
    \% users underweight & 0.34   \\
    \% users normal weight & 29.39 \\
    \% users overweight & 36.28 \\
    \% users obese & 33.99
\end{tabular}
\end{center}
\caption{Descriptive statistics from our dataset of MyFitnessPal users.}
\label{table:1}
\end{table}

\xhdr{The Dataset} We use a dataset of 1.4 million MyFitnessPal users over three years, from August 2014 through April 2017. Users set 2.8 million goals, log 44.6 million weights, and eat 8.8 billion food items. Table \ref{table:1} provides descriptive statistics for our dataset.

In this paper, we focus on weight loss goals and ask the question of which users will ultimately achieve their goal. A user's goal is considered achieved if, at some point in the future after setting the goal, the user logs a weight that is at least as low as that goal. As we are interested in users who showed at least some amount of interest in using the app to track their weight loss progress, we filter to users with at least 7 days between the first and last logged weight of their goal. To remove any minors, extreme outliers, or users who entered likely fake information, we also filter out users who entered a weight over 1000 pounds, an age less than 18 or over 80, and who's weight change goal was more than a 100\% difference from of their current weight. This removed $3,681$ users users.

\section{What weight goals do people choose?}

Performance towards a weight loss goal is, of course, significantly dependant upon the goal's difficulty. This section focuses on goal selection. We aim to validate our dataset by comparing the weight loss goals that users set in MyFitnessPal to the weight loss goals that people set before the age of self-trackers. In particular, we study which goals users choose for themselves; stratifying by demographic information. As mentioned earlier, MyFitnessPal provides no guidance when users are selecting weight targets, which enables an unbiased analysis. While we do note that there is likely a selection bias in terms of who chooses to use MyFitnessPal, towards people who are interested in loosing weight, this bias towards users interested in self-improvement is likely present in many activity tracking apps.

\subsection{Aiming Low: Initial Target Selection}

\begin{figure}
    \centering
    \includegraphics[width=.75\columnwidth]{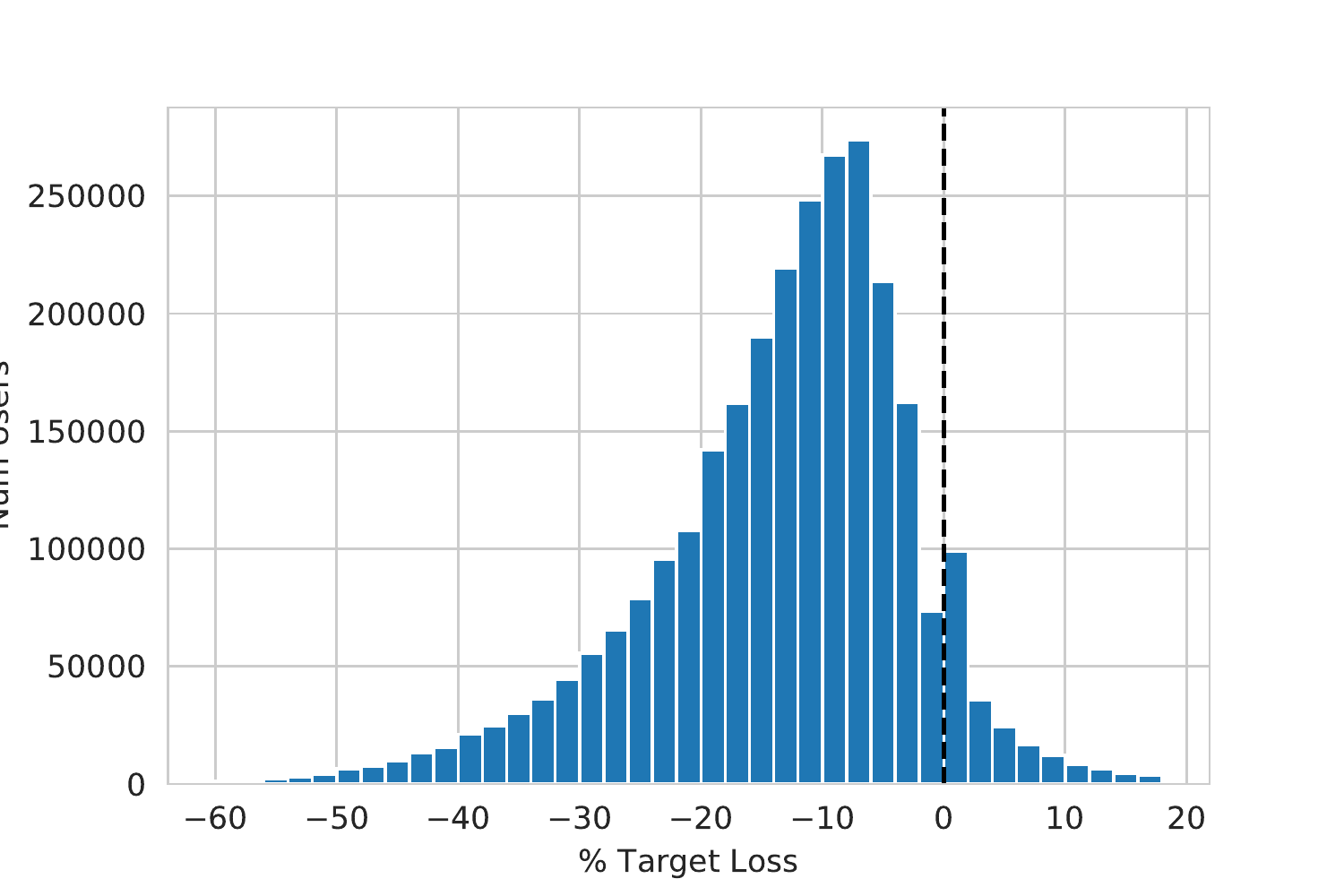}
    \caption{Histogram of weight loss goals as a percent of initial weight. We see a somewhat skewed distribution (skewdness of -0.46), with a peak around 10\% weight loss.}
    \label{fig:hist_target}
\end{figure}

We begin by plotting a histogram of weight loss goals, in terms of percent of current weight, in Figure \ref{fig:hist_target}. We find a somewhat skewed (skewedness = -0.46) distribution with a peak around 10\% weight loss. We also see that a non-trivial portion of users are actually aiming to gain weight. We remove these users for our later analysis, as this is outside the scope of this paper.

\subsection{Declining Ambition: User Age}

\begin{figure}[t!]
\centering
\begin{subfigure}[b]{.75\columnwidth}
   \includegraphics[width=1\linewidth]{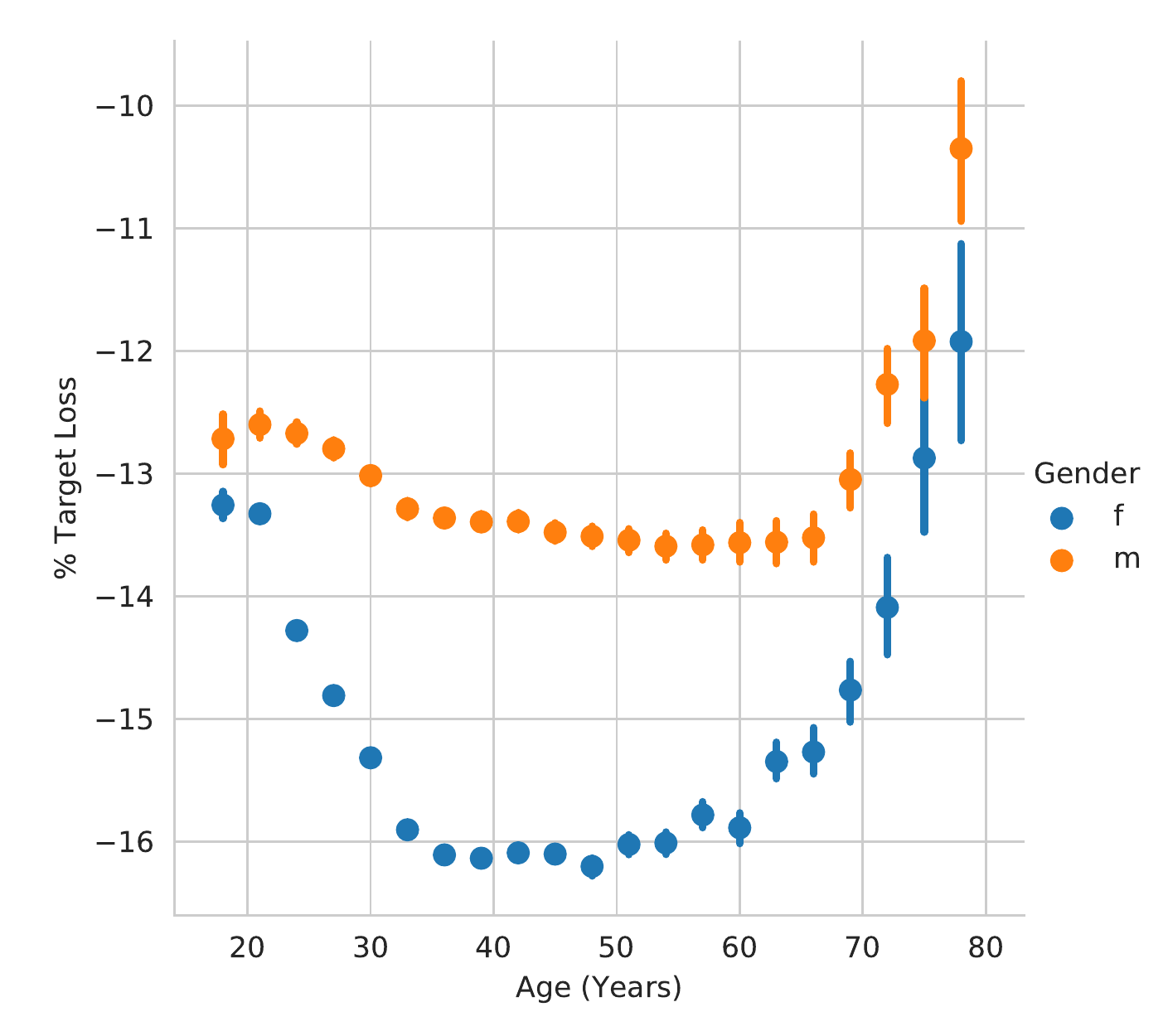}
   \caption{}
   \label{fig:Ng1} 
\end{subfigure}

\begin{subfigure}[b]{.75\columnwidth}
   \includegraphics[width=1\linewidth]{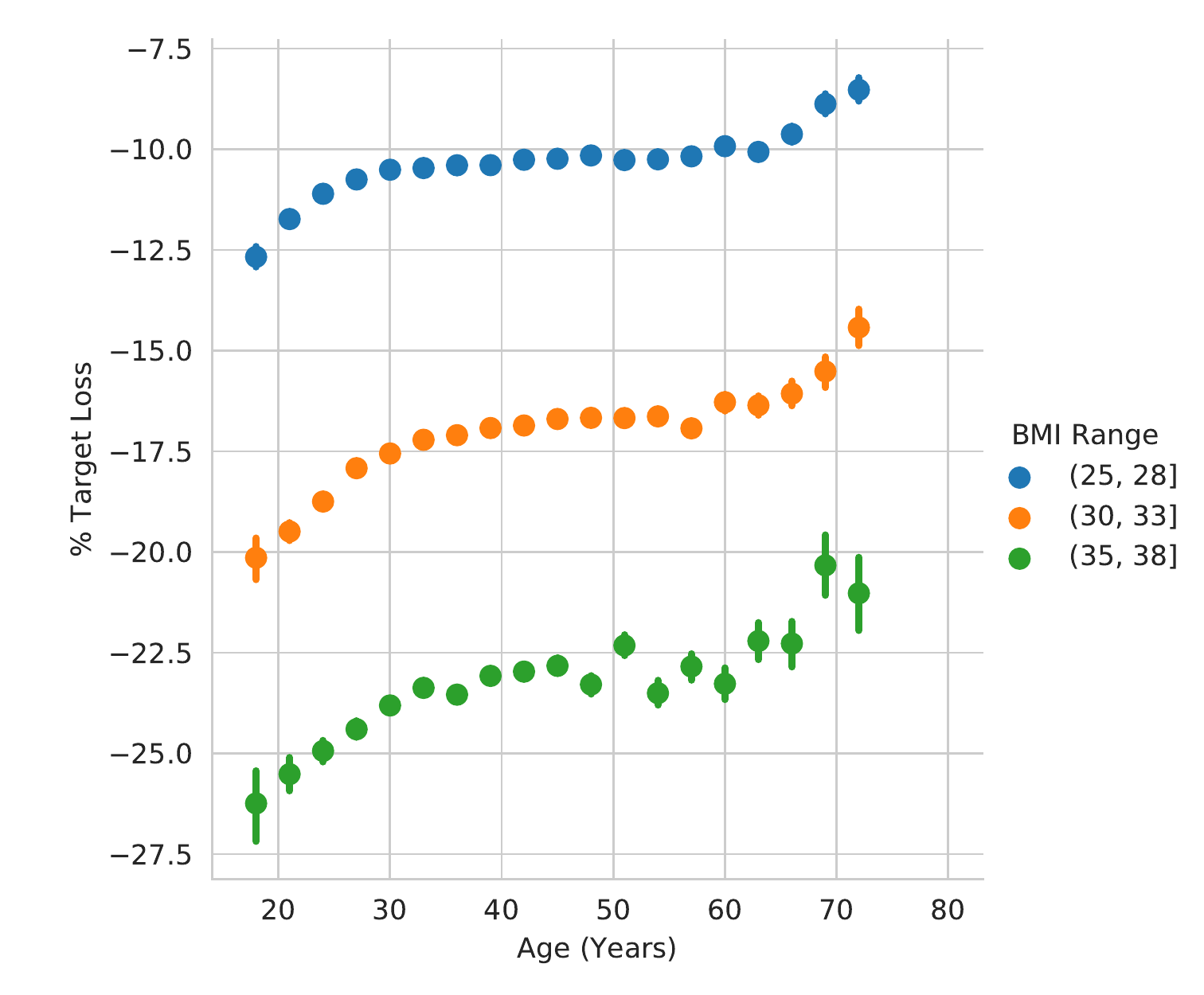}
   \caption{}
   \label{fig:Ng2}
\end{subfigure}
\caption{Weight loss goal as percent of current weight, by age. {\bf (A):} Without stratification by BMI. We find a U-shaped curve in agreement with existing literature. The U is significantly deeper for women than for men. {\bf (B):} With this stratification (results shown only for women), we now see that younger adults set the most ambitious goals, but get less ambitious up until the age of 30, from which point goals stay relatively consistent. Then, around 60, older adults set increasingly less ambitious goals. Error bars in all plots correspond to bootstrapped 95\% confidence intervals (here, they are mostly too small to be visible).}
\end{figure}

In this section, we ask how ambition changes with age. To answer this question, we begin by plotting the amount of weight users aim to lose (as a percent of current weight) by age. We find a U-shaped curve, shown in Figure \ref{fig:Ng1}. For instance, for women between the ages of 18 to 35, users aim to lose an increasingly large percent of their weight, moving from 13.2\% to 15.7\% loss (Mann-Whitney U $p < 0.001$). The number then stays relatively constant around 15.7\% up until the age of 60. However, after age 60, users aim to lose an increasingly smaller amount of weight, with users at age 75 returning approximately to the same goal as those at age 18 (Mann-Whitney U $p < 0.001$).

To validate our results, we can compare them to a 1989 large-scale telephone-based survey of weight loss goals \cite{williamson1992weight}. To our knowledge, this is the most recent such survey, though notably, it occurred long before the existence of activity tracking apps. The survey reports results from 21,109 American households, asking if they were currently aiming to lose weight. If the survey participant said yes, then the study included that goal in their results. This means that both this survey and our behavioral tracking dataset focus on users aiming to lose weight, thus removing the potential for selection bias on that co-variate. The survey reports results separately for men and women.

Similarly to our results, the survey found a U-shaped curve, but with larger magnitudes. For women between ages 18-29, the survey reports a mean of 17.3 percent loss, compared to our 14.3. Between ages 40-49, the survey reports a mean of 18.1, compared to our 16.0. And between ages 70-79, the survey reports a mean of 14.9, compared to our 13.4. The similar shape of their curve to ours helped to validate our results, while at the same time showing that at all ages, users in MyFitnessPal are slightly less ambitious than those in the 1989 survey.

Having validated our dataset, we now ask, how can we explain this U-shaped curve present in both our work and existing work? The U-shape seems to go against what we might expect, given the fact that younger people tend to be less conservative and take more risks than older people \cite{foster1997reasonable}. We note that neither we nor the work we validate our results against have thus far accounted for the fact that age is strongly correlated with whether someone is at a healthy weight, with younger adults more likely to be at a healthy weight. We hypothesize that the U-shaped curve may be due to the fact that younger users will be healthier on average than older users, and thus simply need to lose less to reach a healthy weight. To test this, we take advantage of the unprecedented size of our dataset and stratify our earlier plot by BMI (Body Mass Index, a number based on height and weight used to determine if someone is classified as having a normal weight, overweight, or obese). We do so in Figure \ref{fig:Ng2}. We now find that, when accounting for how healthy a user's existing weight is, we see that younger users actually set significantly more ambitious weight loss goals than older users.

\subsection{Gender Matters: User Gender}

\begin{figure}
    \centering
    \includegraphics[width=.8\columnwidth]{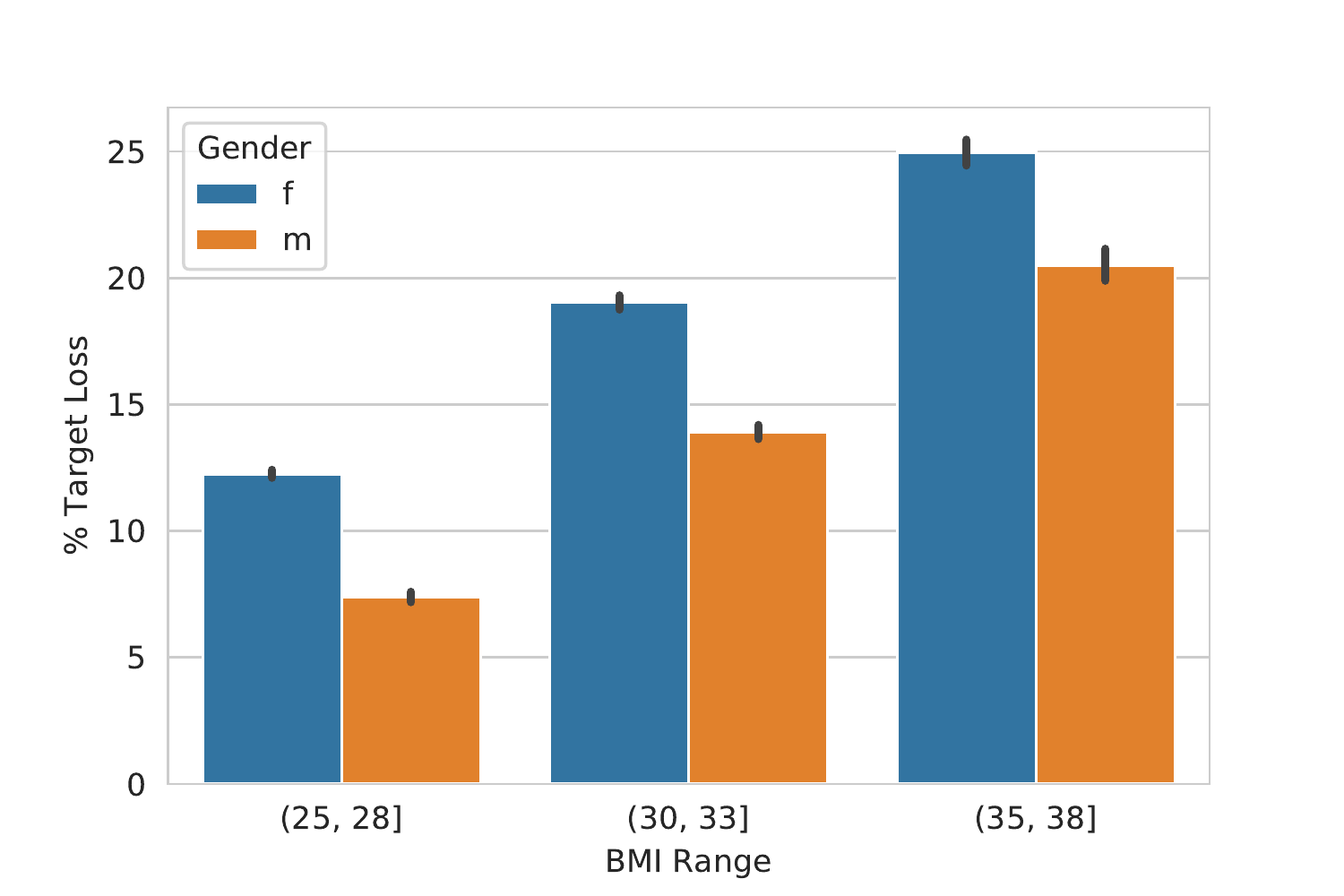}
    \caption{Women set more ambitious goals than men; when controlling for BMI, women aim to lose approximately 5\% more weight.}
    \label{fig:gender_target}
\end{figure}

We now turn to gender differences. Given our findings from the previous section, In Figure \ref{fig:gender_target} we again stratify by BMI to control for selection bias, as men tend to have a higher BMI than women. We find that women set more ambitious goals than men and that, regardless of BMI, this effect size is about a difference of 5\% (Mann-Whitney U $p= < 0.0001$). This is somewhat in agreement with the 1989 survey mentioned above, but with a larger magnitude (they saw a difference of at most 4.1\%).

\section{Do People Reach Their Goals?}
Next, we explore goal achievement rates along with the amount of time taken for goal achievement. We find that overall, 18.2\% of weight loss goals are met. This is remarkably close to an earlier finding from 2005 that approximately 20\% of people overweight individuals are successful at long-term weight loss \cite{wing2005long}.

\subsection{How many people reach their goals?}

\begin{figure}
    \centering
    \includegraphics[width=.8\columnwidth]{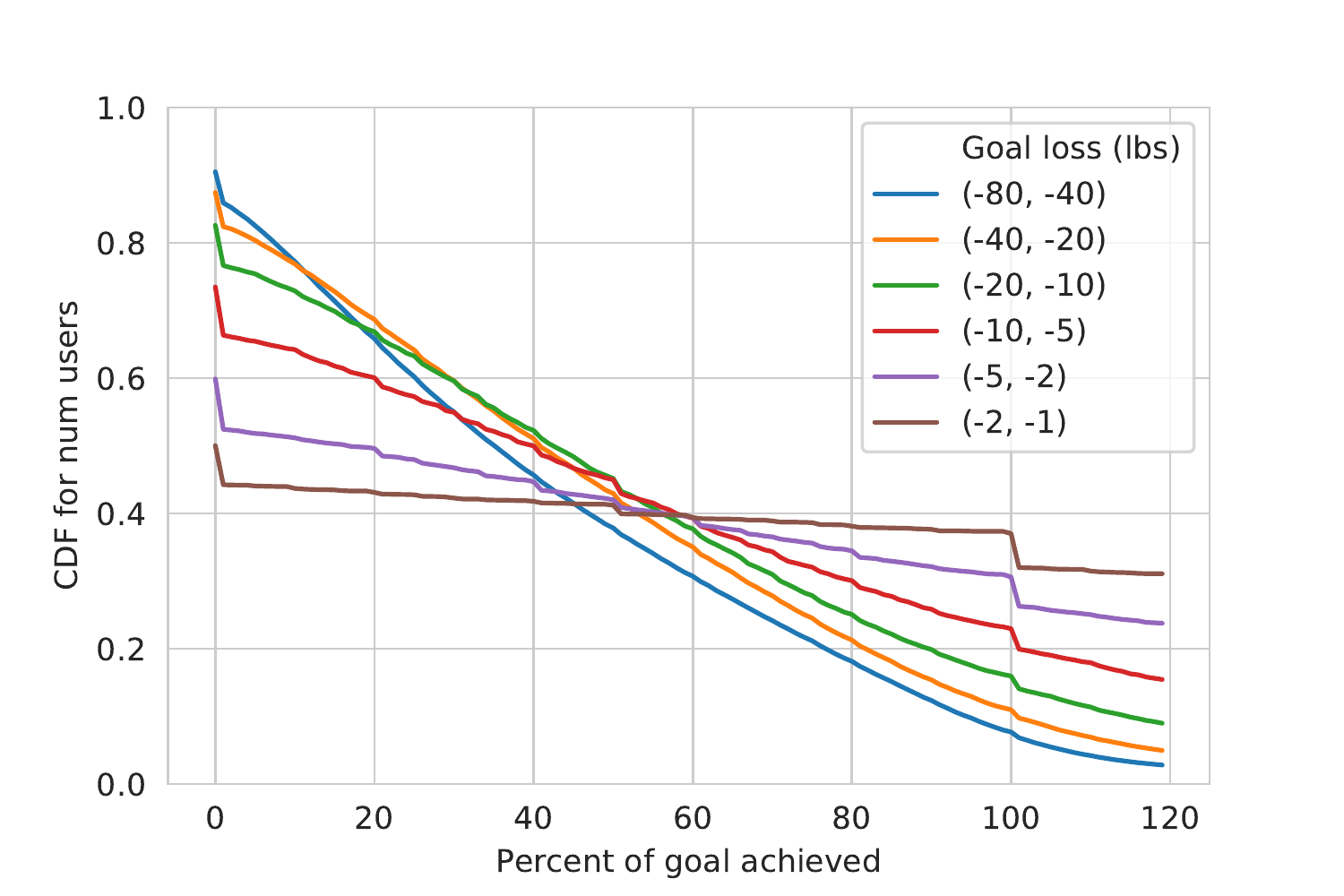}
    \caption{CDF of goal achievement, stratified by goal difficulty. $X$-axis indicates percent of goal achieved while $y$-axis indicates the portion of users who have achieved at least that amount of their goal. We find that across the board, goals are unlikely to be achieved. Only 30\% of the easiest goals are ever reached, while less than 5\% of the most difficult goals are ever reached.}
    \label{fig:ga}
\end{figure}

Clearly, goal difficulty should impact goal achievement. We hypothesize that more difficult goals are less likely to be achieved. Therefore, we start by breaking down target achievement as a function of goal size. Figure \ref{fig:ga} shows a cumulative distribution function (CDF) with a different line for target size buckets. The $x$-axis indicates percentage of target reached ($x=100$ indicates that a target has been achieved). We find that target achievement is relatively low, and decreases as targets become more ambitious. People with the easiest goals (between 1 and 2 percent of their initial weight) are most likely to achieve their targets, with nearly 30\% of these users reaching at least 100\% of their initial weight loss goal. Of users with the most difficult goals (between 40 and 60 percent of their initial weight), only 10\% are likely to achieve their target. 

The $x$-axis of the above plot begins at zero, indicating that a user's weight has not changed. We note that the plotted $y$-value of our CDF does not begin at 1.0 for any of the goals; this is because a large portion of users end up gaining weight, putting their goal achievement progress to the left of our $x$-axis starting point. Interestingly, despite that fact that users who set easier goals are more likely to achieve them, we also find that users who set easier goals are more likely to go in the opposite direction and gain weight. For instance we see that only 40\% of users with the easiest goals lose weight, while 90\% of users with the hardest goals end up losing weight. This finding highlights the importance of not setting a goal that is ``too easy'' for a user.

We also notice a blip at 100 percent achievement, indicating that while many users either change their target or stop using the app once they've reached their goal, others continue to lose weight even after achieving their target.

We also find that men are more likely to meet their weight loss goals, even when accounting for the fact that men set easier goals: for instance, men who aim to lose 5-10\% of their weight succeed 28\% of the time, while women who aim to lose 5-10\% of their weight succeed 20\% of the time.

\subsection{How long does it take to reach your target?}

\begin{figure}
    \centering
    \includegraphics[width=.8\columnwidth]{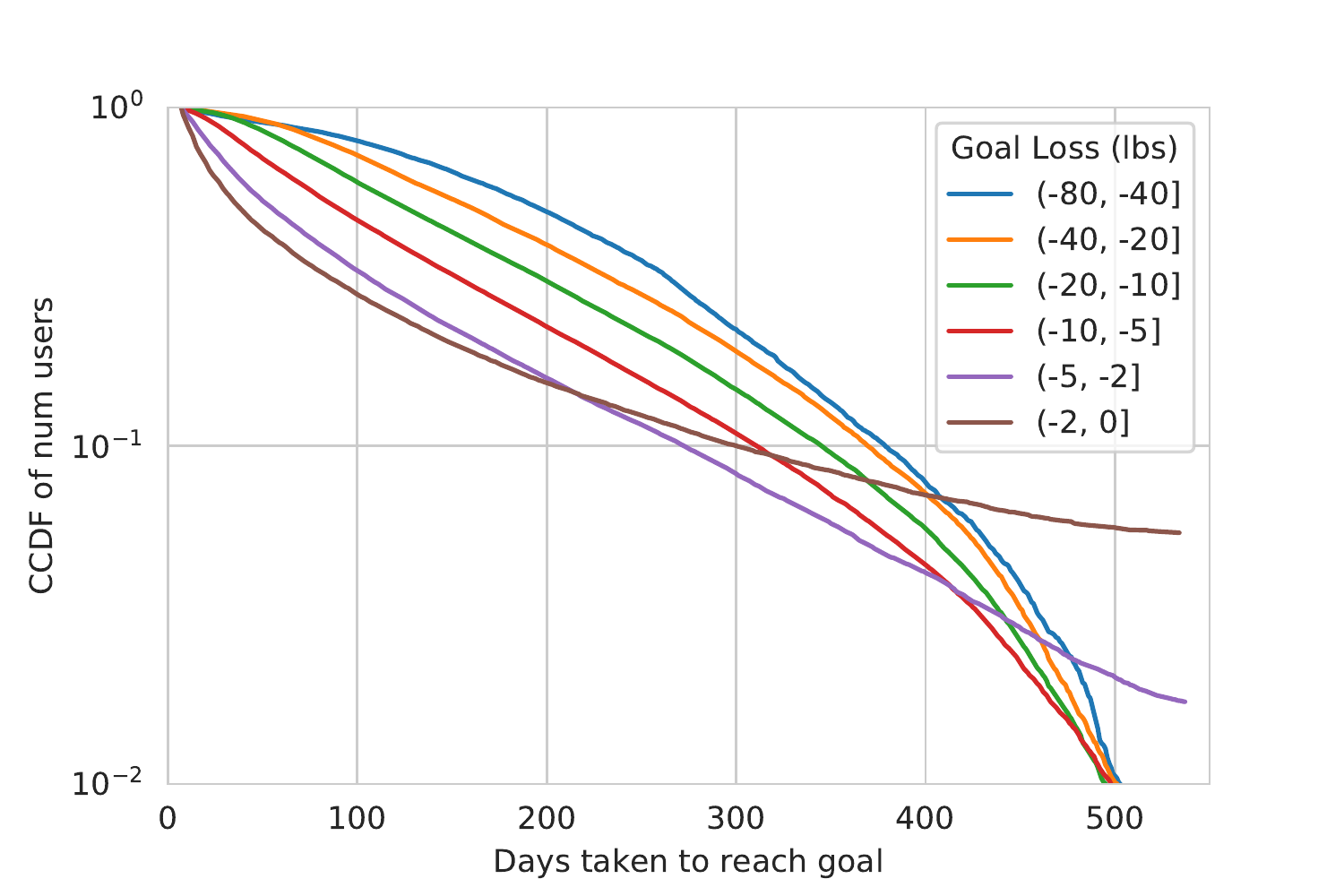}
    \caption{CCDF of days taken to achieve a goal, stratified by goal difficulty. More difficult goals take more time to achieve.
    }
    \label{fig:gt}
\end{figure}

Having established how many people reach their targets, we now ask how long it takes them to do so. In Figure \ref{fig:gt}, we show a CCDF of time taken to achieve a goal as function of goal difficulty. We find that more difficult goals take longer to achieve. For the easiest goals, half of all users who meet their goal do so within 100 days. For the most difficult goals, this number is closer to 300 days.

\section{Early Warning Signs}
Having seen that few users ever meet their goals, we now ask what are the early behavioral factors that indicate a user will or won't meet their goal? We ground our analysis in the existing body of literature on short and long term weight loss. We work to both answer open questions in the literature and to determine how existing findings apply specifically to early detection in weight loss target achievement.

We focus specifically on the first seven days after the first weight of a goal has been logged. We find that even for difficult, long-term goals such as losing twenty pounds, behavioral differences observed during just the first seven days are indicative of large changes -- up to double -- the propensity to reach that goal.

\subsection{Initial Weight Loss Patterns}

\begin{figure}
\centering
\begin{subfigure}[b]{.8\columnwidth}
   \includegraphics[width=1\linewidth]{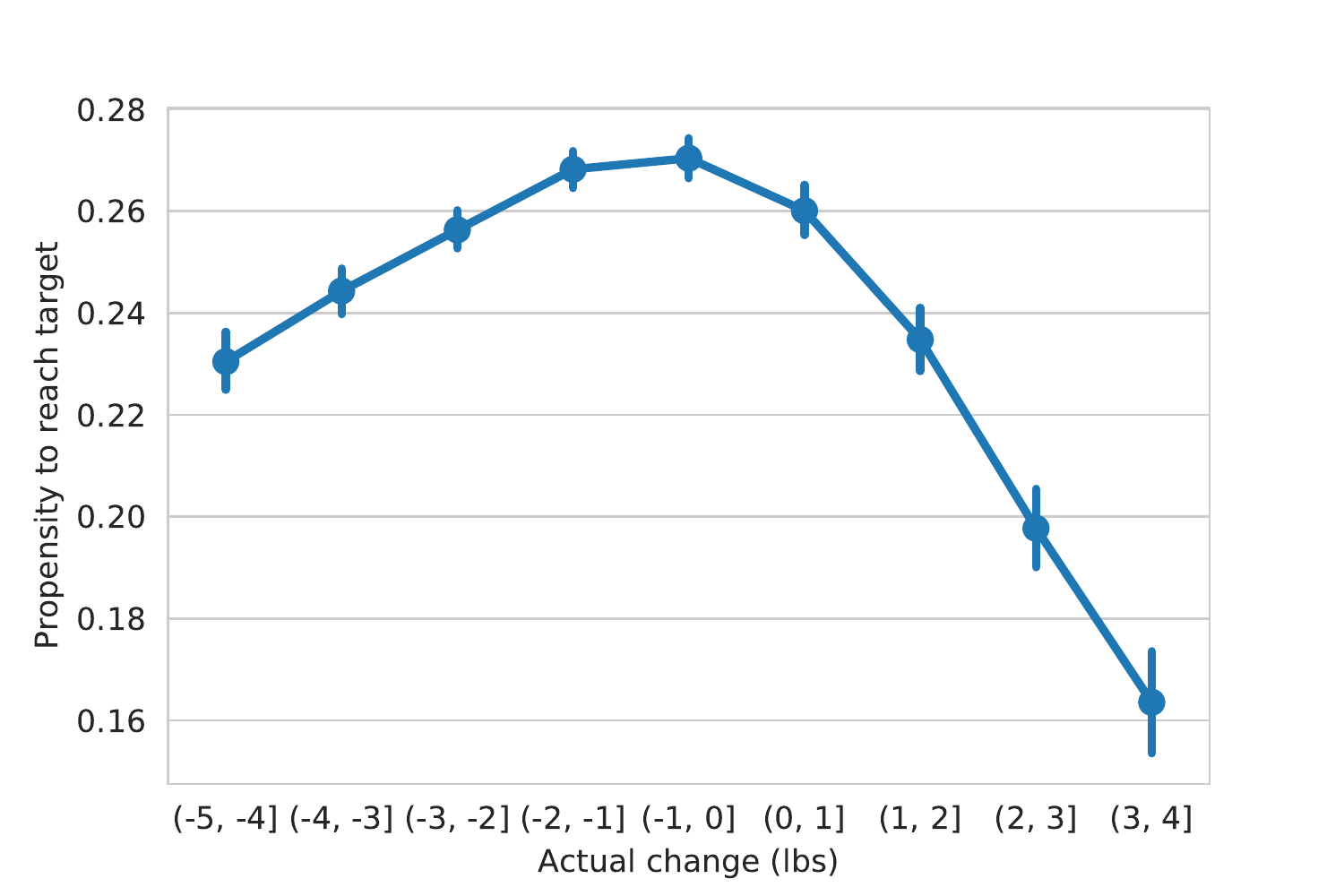}
   \caption{}
   \label{fig:1weekprogu} 
\end{subfigure}

\begin{subfigure}[b]{.8\columnwidth}
   \includegraphics[width=1\linewidth]{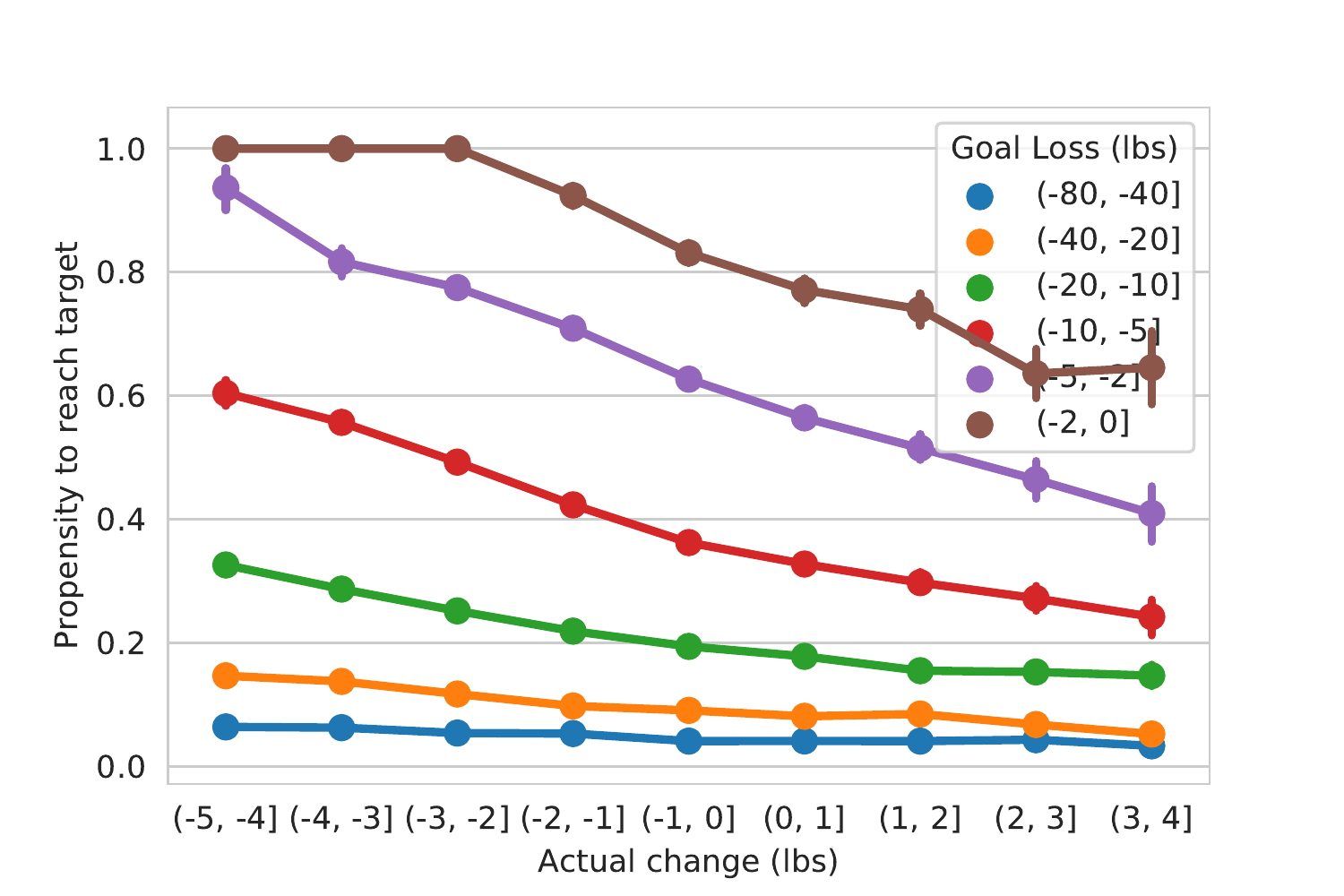}
   \caption{}
   \label{fig:1weekprog}
\end{subfigure}
\caption{Propensity to meet weight loss goal by weight loss over the first week. {\bf (A):} Without stratification by goal difficulty, we find a U-shaped curve in agreement with existing literature suggesting that ``too much'' early weight loss is unsustainable and leads to worse longer-term outcomes. {\bf (B):} With stratification by goal difficulty, we now see that people who lose more weight in the first week are far more likely to reach their goals. Contrary to what medical literature might suggest, there is no such thing as ``too much'' unsustainable early weight loss which ultimately leads to worse results.}
\end{figure}

\xhdr{Too much too fast?} Initial weight loss has been identified as a predictor for later weight loss, and also for weight loss maintenance \cite{elfhag2005succeeds}. Some studies have found that the greater the initial weight loss, the better the subsequent outcome \cite{astrup2000lessons}. However, other work has found that greater initial weight loss actually predicts future weight gain, explained by the fact that overly quick weight loss is considered unsustainable and clinically unhealthy \cite{mcguire1999predicts}.

Goal-setting theory suggests that early progress towards a goal could serve to provide future motivation, while slow initial progress may cause frustration and ultimately lead to an abandonment of the goal \cite{locke2002building}. This taken together with the clinical findings discussed above, make it uncertain as to whether there may be such a thing as ``too much'' initial progress towards a weight loss goal.

To answer this question, first plot propensity to meet weight loss goal by weight change over the first seven days. Figure \ref{fig:1weekprogu} shows that we find a U-shaped curve suggested by prior weight loss maintenance literature. On the right, we see that as people gain more weight during the first week, that are less likely to meet their weight loss goals. On the left we see that, as people lose more weight, they are also less likely to meet their goals. We note that the downward trend seems to start around -2 lbs per week, which is exactly what prior literature would suggested: 2 lbs per week of weight loss is often considered the maximum healthy amount \cite{hollis2008weight}.

However, we hypothesize that there may be a different reason for this U-shape curve than the idea that ``too much'' early weight loss is unsustainable and leads to weight gains, as suggested by prior literature. We hypothesize that this U-shaped curve may be due to the fact that people who lose less weight during the first week have also set easier goals on average. To test this, we can stratify our earlier chart by goal difficulty. We show this result in Figure \ref{fig:1weekprog}. We now find that there is in fact no amount of initial weight loss which reduces a user's propensity to meet their weight loss goal. Further, see a large difference in propensity to meet goals based on initial weight loss: for instance, a user who loses between 4-5 pounds over the first two weeks of their 20-40 pound loss goal is twice as likely meet that goal as one who loses between 1 and 2 pounds (Mann-Whitney U $p < 0.001$). This finding suggests that follow-up studies to highly cited clinical weight loss papers may be warranted, to determine whether the idea of ``too much early weight loss'' is due to confounding by motivational reasons rather than physical limitations around unsustainable weight loss

\subsection{Self-monitoring}

\begin{figure}
    \centering
    \includegraphics[width=.8\columnwidth]{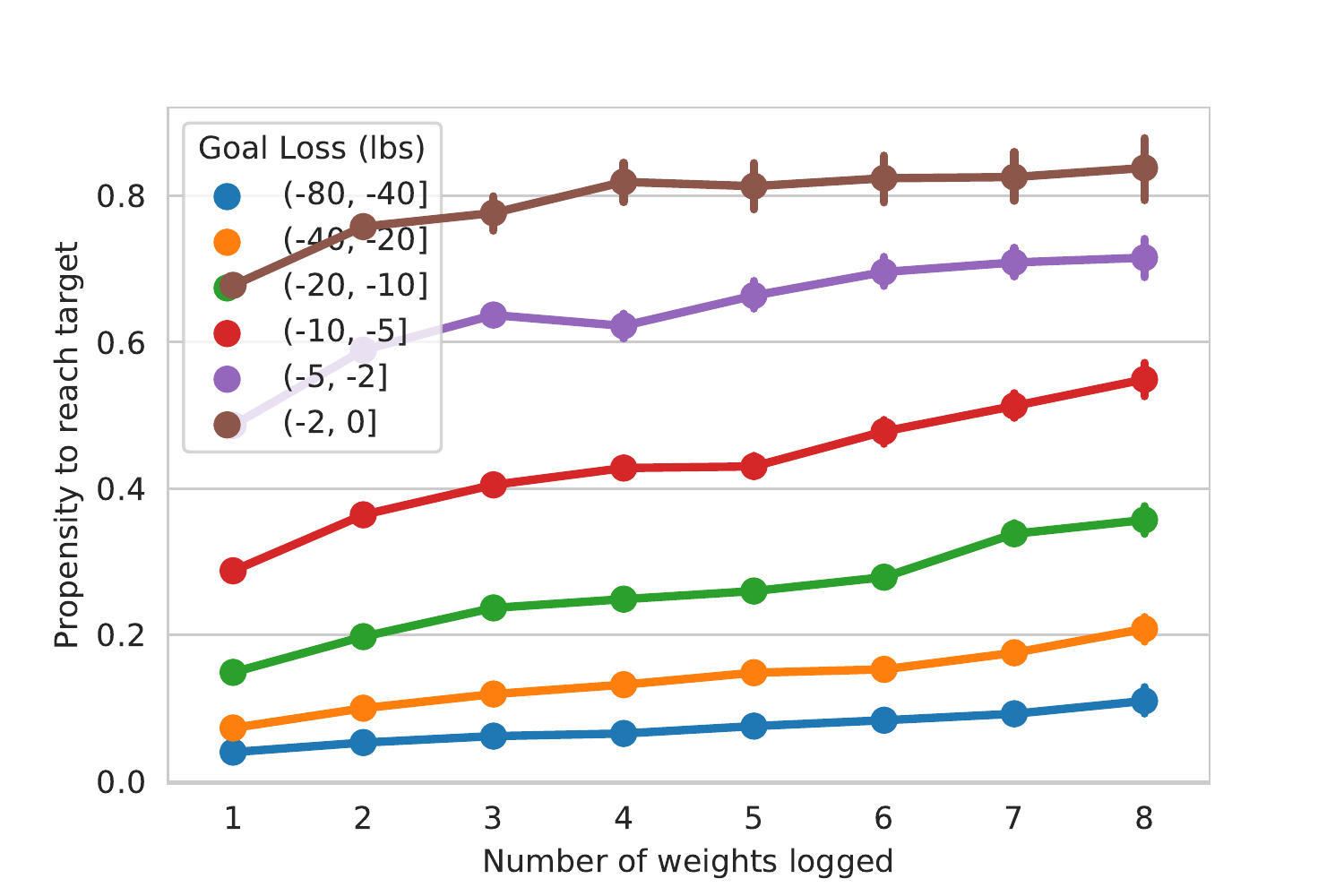}
    \caption{Propensity to achieve goal by number of days logged a weight over the first seven days, stratified by goal difficulty. We see that across all difficulty of goals, people who log weights more frequently over the first week are more likely to achieve their goal.}
    \label{fig:selfmonitor}
\end{figure}

Self-monitoring means observing oneself and one's behaviour. Indeed, this is the core purpose of an activity tracker. Most activity tracking apps today rely on their users manually choosing log many important behaviors. While step count is typically logged automatically, workouts, food intake, and weights often involve some amount of manual input.

Regular self-monitoring of weight has been linked to long-term weight loss maintenance, which suggests that it may also be important for meeting weight loss goals \cite{burke2011self}. In Figure \ref{fig:selfmonitor}, we plot the number of days that a user logged their weight over the first 7 days of a goal. We see that, for all difficulty of weight goals, people who log more regularly over the first seven days are significantly more likely to ultimately meet their goal. For a goal between 40-80 lbs, users who log a weight all 7 days of their first week are more than twice as likely to ultimately meet their goal than those who log only once, moving from 10\% to 20\% likelihood (Mann-Whitney U $p < 0.001 $). Similarly, for a goal between 10 and 20 lbs, we see a change between logging one weight and seven weights from 15\% up to 35\% (Mann-Whitney U $p < 0.001 $).

\subsection{Dietary Intake}

\begin{figure*}[h!]
  \centering
  \begin{subfigure}[b]{0.3\textwidth}
    \includegraphics[width=\textwidth]{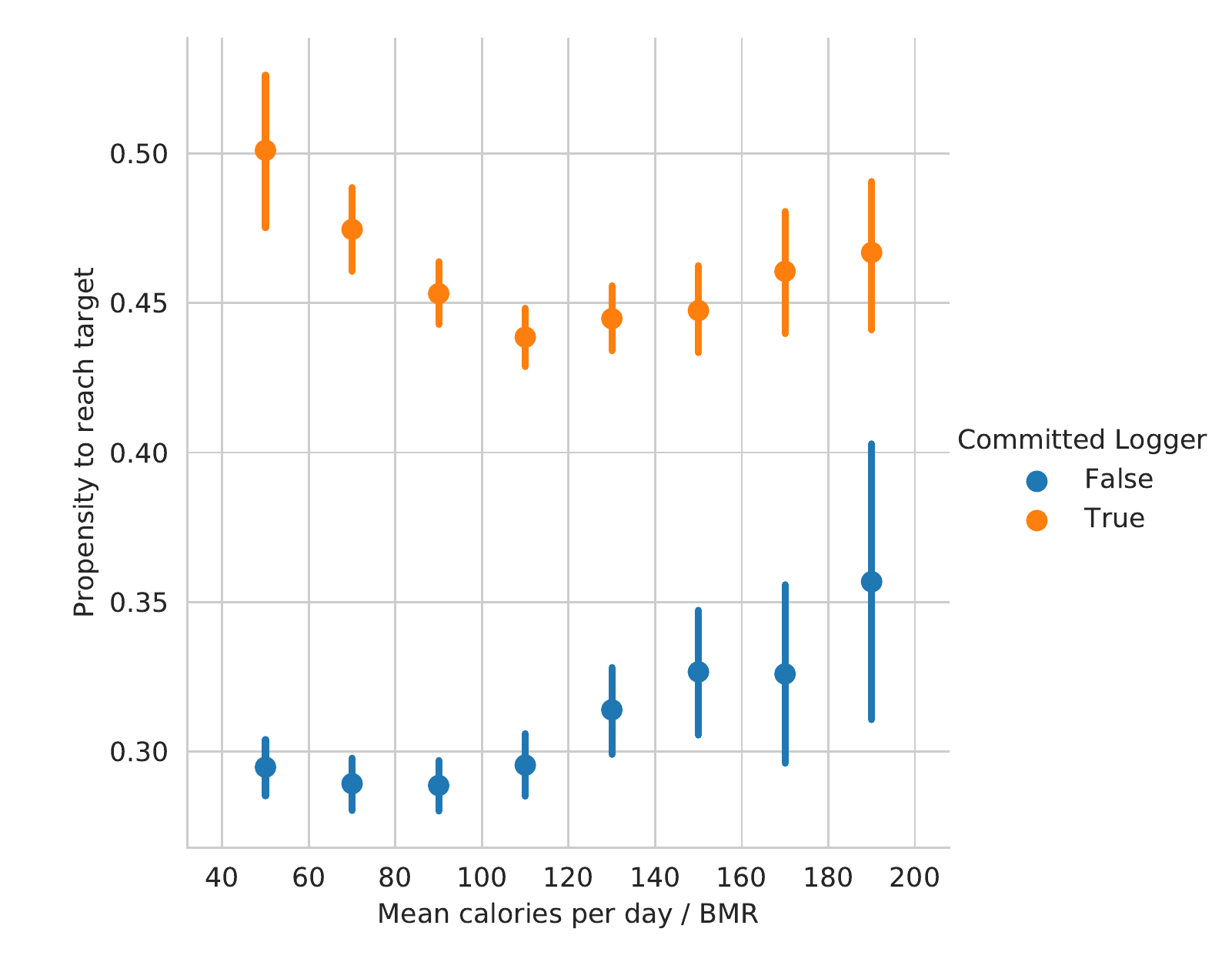}
    \caption{5-10 lb goal loss.}
  \end{subfigure}
  \begin{subfigure}[b]{0.3\textwidth}
    \includegraphics[width=\textwidth]{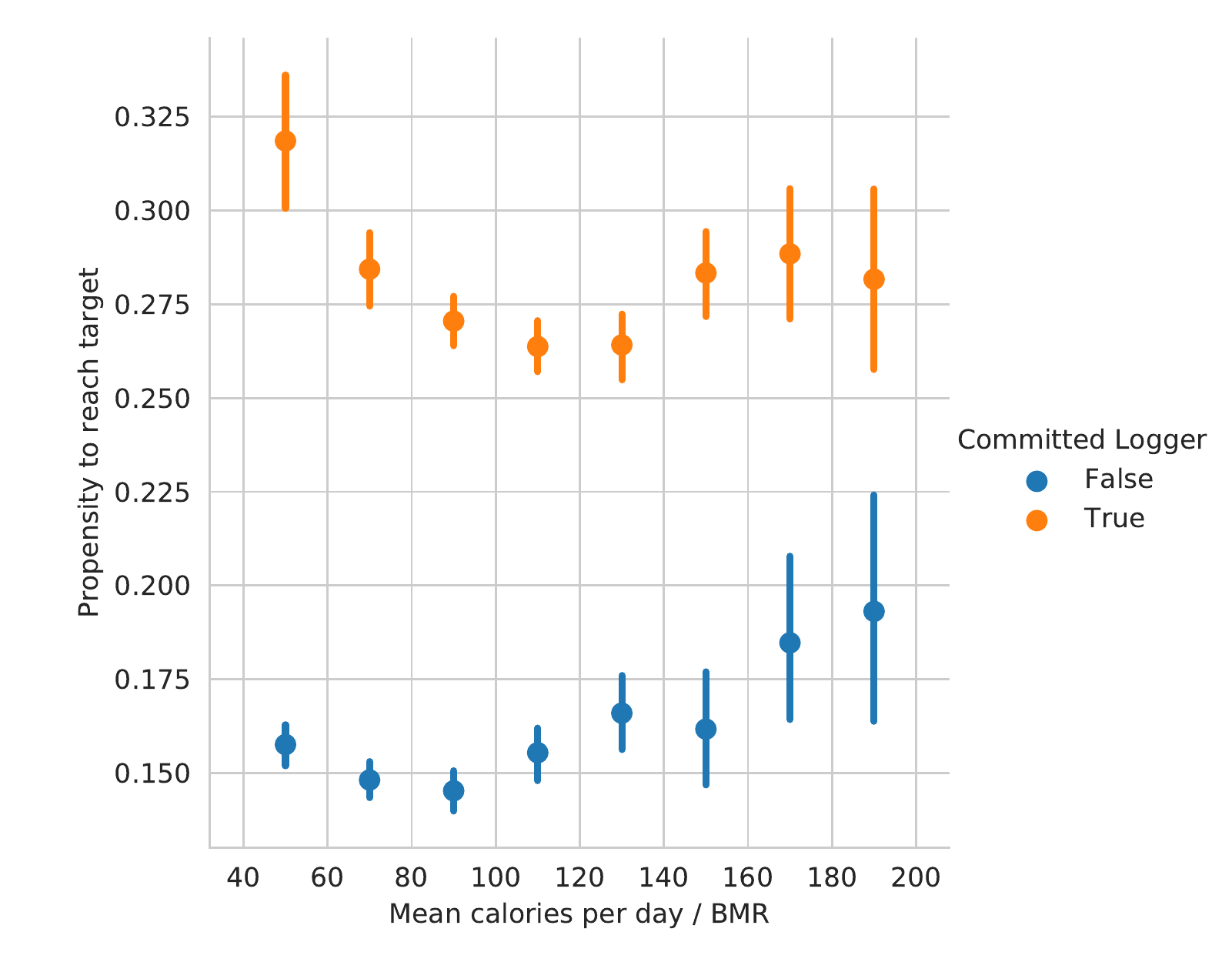}
    \caption{10-20 lb goal loss.}
  \end{subfigure}
    \begin{subfigure}[b]{0.3\textwidth}
    \includegraphics[width=\textwidth]{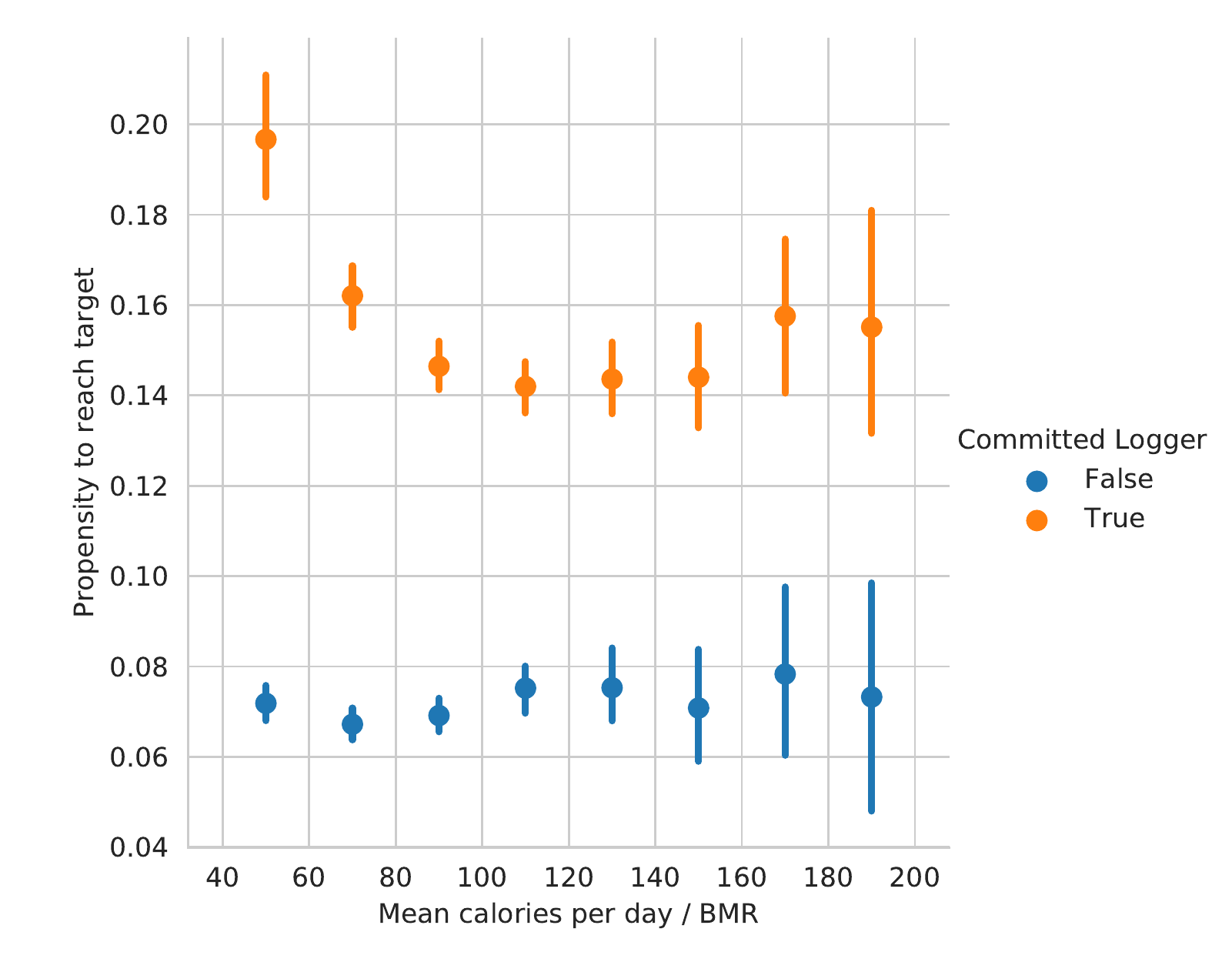}
    \caption{20-40 lb goal loss.}
  \end{subfigure}
  \caption{Committed food loggers during the first 7 days of a goal are much more likely to meet their weight loss goals than less committed loggers. Further, committed loggers who log more calories are less likely to meet their goal, while uncommitted loggers who log more calories are more likely to meet their goal. }
  \label{fig:calsgoalloss}
\end{figure*}

One of the core features of MyFitnessPal is the ability to log food intake. Obviously, food intake is strongly associated with weight loss. Longer-term weight loss and maintenance has been linked to eating smaller portions and higher quality nutrients \cite{elfhag2005succeeds}.

In MyFitnessPal, users log individual food items, where each item has an associated calorie amount (typically populated by MyFitnessPal's food database). We hypothesize that users might fall into two categories: uncommitted loggers, who likely aren't particularly motivated to log their food and don't log everything they eat, and committed loggers, who are motivated to log what they eat. We assign users to groups based on total days logged over the first 7 days: users who logged at least one food item all 7 days are considered a committed logger, and users who did not are considered an uncommitted logger. How likely are these groups to meet their weight loss goals? 

Figure \ref{fig:calsgoalloss} shows propensity to meet a weight loss goal by mean calories per day over the first 7 days, normalized by each user's Basal Metabolic Rate (BMR, the amount of calories an individual consumes per day at rest), for both committed and uncommitted food loggers. To avoid confounding by goal difficulty, each chart contains only a specific range of goal weight losses. We find that, regardless of amount calories logged, committed food loggers are far more likely to meet their weight loss goals. We also see that, as committed loggers log more calories, they are less likely to meet their goals, which is what we would expect given that eating more calories results in worse weight loss. However, this effect only lasts up until around 110\% BMR calories logged, after which more calories logged actually seems to correspond with a higher propensity to meet the goal (though we note that confidence intervals all overlap). This is likely explained by the fact that some of these users are not actually eating more, but rather are extreme loggers who log all food that they eat. 

For uncommitted loggers, we see a very slight curve for which more eating indicates a lower propensity to meet a goal, followed by a higher propensity. The difference between the top of the curve and the bottom of the curve is extremely small, which is likely because for these uncommitted loggers, more food loggers mostly is indicative of an increasingly higher commitment to logging, rather than more eating. Overall, we find that logging more calories leads to the expected reduction in propensity to meet a weight loss goal, but only for committed food loggers.

\subsection{Motivation}

\begin{table}[]
\begin{tabular}{|l|l|l|l|}
\hline
\textbf{1. Kids role model} & \textbf{2. Lifestyle} & \textbf{3. Health} & \textbf{4. Clothes} \\ \hline
good                      & live               & shape           & fit               \\ \hline
example                   & life               & health          & clothes           \\ \hline
set                       & long               & weight          & old               \\ \hline
family                    & longer             & family          & able              \\ \hline
children                  & enjoy              & need            & jeans             \\ \hline
kids                      & happy              & tired           & wear              \\ \hline
daughter                  & family             & time            & energy            \\ \hline
energy                    & able               & lose            & cute              \\ \hline
role                      & kids               & kids            & size              \\ \hline
model                     & time               & energy          & comfortable       \\ \hline
\end{tabular}
\caption{The four topics discovered from NMF topic model trained on weight loss goal reasons supplied by users when setting their goal. Each topic lists the top ten words associated with it.}
\label{table:topics}
\end{table}

\begin{figure}
    \centering
    \includegraphics[width=.75\columnwidth]{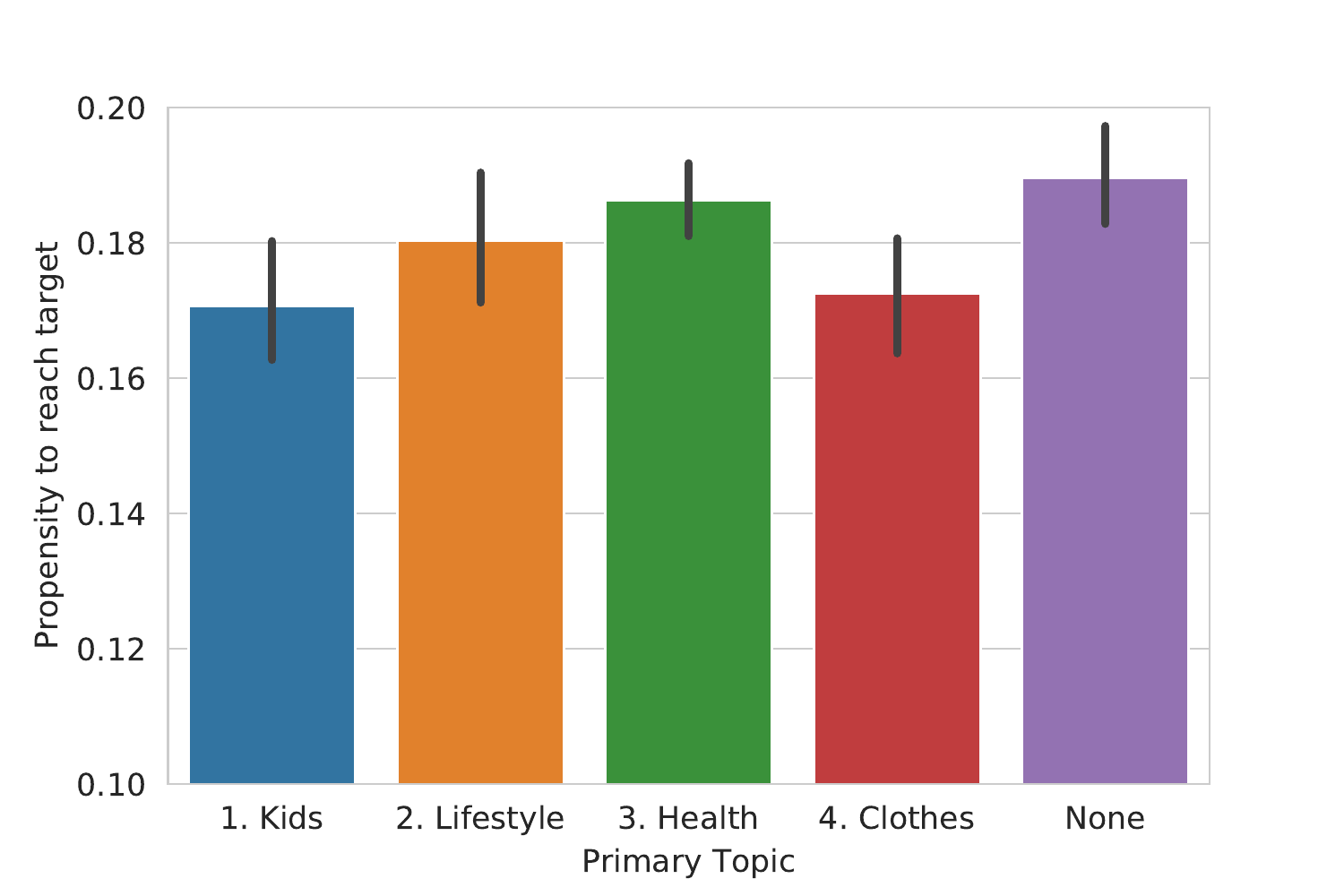}
    \caption{Propensity to achieve goal by motivation. Motivations derived from learning a topic model over free-text goal reasons provided by users and then assigning each goal a primary topic.}
    \label{fig:topicresults}
\end{figure}

When a user sets their target, MyFitnessPal optionally allowed users to fill in a free text form which asked why they were selecting their goal. We call this a ``weight loss reason''. Attitude has been found to be important in weight loss maintenance \cite{elfhag2005succeeds}, which suggests it may also be important in meeting weight loss goals. We hypothesize that the weight loss reason provided by a user may provide a glimpse into their mindset and attitude towards setting and meeting their goal. In this section, we examine these weight loss reasons to determine whether they are indicative of propensity to meet a weight loss goal.

After filtering to users who choose to use this field, we end up with 56,014 goals associated with a reason. We first provide a few sample reasons to give a sense of the type of reasons provided by MyFitnessPal users (note that for anonymization, we slightly change the exact text while retaining the meaning). Many are somewhat generic, with users writing that they are setting their goal ``To feel better, and have more energy!'' and ``I would like to get in shape and get back to the weight where I felt comfortable with my body.'' However, some reasons were quite specific to the user's personal situation, such as ``My dad's adopted and doesn't know his family history, but he had a heart attack at age 34 and a RI before age 46. I already have heart issues (which admittedly are well controlled with meds, but still.....''

The choice to fill in this field seems likely to be related to propensity to self monitor and commitment to the app which, as discussed earlier, are linked to a higher propensity to achieve one's goal. Surprisingly, we find that users who fill in a weight loss reason are less likely to meet their goals (20.5\% vs 17.6\%) (Mann-Whitney U $p < 0.001 $). However, this difference can likely be explained by the fact that people who supply a weight loss reason aim to lose nearly 50\% more weight (33.8 lbs vs 24.6 lbs, Mann-Whitney U $p < 0.001 $) and, as we showed earlier, more difficult goals are less likely to be met.

We now ask: what is different about the reasons given by people who meet their goals versus the people who don't? We first find that people who meet their goals use 1.4 fewer words (Mann Whitney U $p < 0.0001$) than those who do meet their goals, at 19.2 words compared to 20.6 words. This is surprising, as one might expect that a longer reason indicates more commitment to the app or more thought-out reasons for losing weight. But how do these reasons differ in terms of the content's subject matter?

\xhdr{Topic Models} Studies have found that people who are motivated to lose weight by a desire for greater self-confidence or superficial considerations such as appearance (e.g. a  ``healthy narcissism'') are more successful at maintaining weight loss than those who are primarily motivated by medical concerns or from pressure by those around them \cite{elfhag2005succeeds}. To determine if we see a similar effect in propensity to meet weight loss goals, we would like to map each weight loss reason to an easily interpretable category. To do this, we propose a novel application of topic modeling algorithms. 

Topic models are an unsupervised learning approach to automatically infer interesting patterns in large text corpora \cite{paul2014discovering}. We use this exploratory approach with the goal of understanding what topics are commonly discussed in goal reasons provided by users, with the aim of understanding how they differ in goals which are or are not achieved. Topic modeling has previously been used to extract health-related topics from Twitter \cite{paul2014discovering,ghosh2013we}.

We build our topic model using non-negative matrix factorization (NMF) on extracted term frequency-inverse document frequency (tf-idf) features from our corpus, which helps to select the most useful terms. We use the ``English'' stopwords list from scikit-learn \cite{scikit-learn} along with a few words we added manually: want, look, feel, better, healthy, healthier. We use these as stopwords because they were present in far too many articles, to the point that they dominated every topic generated, and in this situation a typical approach to improve topic models is to remove such words. We experiment with selecting a number of topics between k=3 and 20, ultimately selecting k=4 after finding them the most clearly interpretable and differentiated.

Table \ref{table:topics} presents the top 10 terms from each of the four discovered topics, along with our manually assigned topic labels. We assign each goal reason a ``primary topic'' by selecting the topic with the highest associated score from our NMF model, with a minimum required score of 0.01 (22.4\% of goals did not meet this threshold for any of the four topics). We find the topics to be:

\begin{enumerate}
  \item \textbf{Role model for kids:} users with a reason in this topic discussed how they wished to set a better example for their children. Primary topic for 12.4\% of users.
  \item \textbf{Lifestyle:} users wanted to live longer, happier lives so that they'd have more time to spend with their family. Primary topic for 12.8\% of users.
  \item \textbf{Health:} users were concerned about their health and wanted to get in better shape. Primary topic for 36.9\% of users.
  \item \textbf{Clothes:} reasons in this topic discussed wanting to fit into older clothes that they used to be able to fit in to, or being able purchase and fit into new clothes that they found attractive. Primary topic for 15.4\% of users.
\end{enumerate}

Now that we have learned topics for our goal reasons, we can examine which reasons reflect a higher propensity to meet one's goal. To do this. In Figure \ref{fig:topicresults}, we plot propensity to meet goals by primary topic. We first note that, unlike in prior sections of this paper, we are unable to stratify by goal difficulty due to the smaller size of this dataset. This means that some results may be explained by the fact that, for instance, people who's primary motivation is fitting in to clothes set slightly easier goals (~2 lbs) than those who primary motivation is health. That said, we do find that people primarily motivated by health are more likely to meet their goals than people primarily motivated by clothing, which seemingly goes against what the prior work cited above would suggest.

\section{Predicting goal achievement}
Next, we build on insights from previous sections to predict the likelihood that a user will ever achieve their goal after observing their behavior over the first 7 days.

\subsection{Features Used For Learning}
To illustrate the predictive power of the different feature sets reported earlier in this paper, we define a series of models, each with a different feature set corresponding to one of the goal-related behaviors. We focus on six types of features:

\begin{enumerate}
  \item \textbf{Weight loss goal:} In Section 4, we saw propensity to meet a weight loss goal varied significantly with the difficulty of the goal, with easier goals more likely to be met.
  \item \textbf{Demographics:} we found that propensity to meet a goal also varied with the user's age and gender, with men and younger adults more likely to meet their goals.
  \item \textbf{Initial weight loss:} In Section 5.1, we saw that people who lost more weight over the first 7 days were more likely to meet their goal.
  \item \textbf{Self-monitoring:} In Section 5.2, we saw that users who log their weights more frequently over the first 7 days are more likely to meet their goals.
  \item \textbf{Calories logged and frequency of logging:} In Section 5.3, we saw that logging more calories indicates a lower likelihood of meeting a goal, but only for committed loggers.
  \item \textbf{Motivation:} In Section 5.4, we analyzed free-text motivation provided by users. We use word-level TF-IDF results (for the top 5000 words) as features in this model. This is the same methodology used to train our topic models discussed in Section 5.4.
\end{enumerate}

\subsection{Experimental Setup}
We aim to predict the likelihood that a user will ever achieve their weight loss goal, after observing their behavior only in the first 7 days. As in Section 6, we filter to users for who there is at least 7 days between the start of their goal and the last weight logged of their goal (though we do not require a minimum number of weight logged during this period). We also remove all users who have already met their goal within the first several days, as it would not be a fair task to predict goal achievement after a user has already achieved their goal. We note that, as shown in Figure 6, most users take far longer than 7 days to meet their goals. We also remove users who didn't last at least 7 days to avoid users who download the app as a curiosity and use it very briefly without ever truly attempting to meet their goal. Further, we filter to users who use the features of the app that we discuss in this paper at least once: specifically, users must log at least one food item, and also must fill in the free-text motivation field when setting their goal. After applying these filters, we arrive at a dataset containing 33,642 goals. (This large reduction from our overall dataset is primarily due to the fact that most users chose not to provide a free-text weight loss reason, as the app provided no motivation for them to do so. Without that restriction, we would have over 900,000 goals in our prediction dataset, and we find that ROC AUC would be only slightly reduced).

We experiment with several classification models, including Linear Regression, Gradient Boosted Trees, Support Vector Machines, and Random Forests. We find that Random Forest models are the most effective for our task, and so we report results from only from those models. Because of the unbalanced dataset (only 18.6\% of users achieve their goal) and the trade-off between true and false positive rate associated with prediction, we choose to compare models using the area under the receiver operating characteristic (ROC) curve (AUC) which is equal to the probability that a classifier will rank a randomly chosen positive instance higher than a randomly chosen negative one. Thus, a random baseline will score 50\% on ROC AUC. We use a 10-fold cross-validation for estimation. We standardize all features to have zero mean and unit variance.

\subsection{Results}
\begin{figure}
    \centering
    \includegraphics[width=\columnwidth]{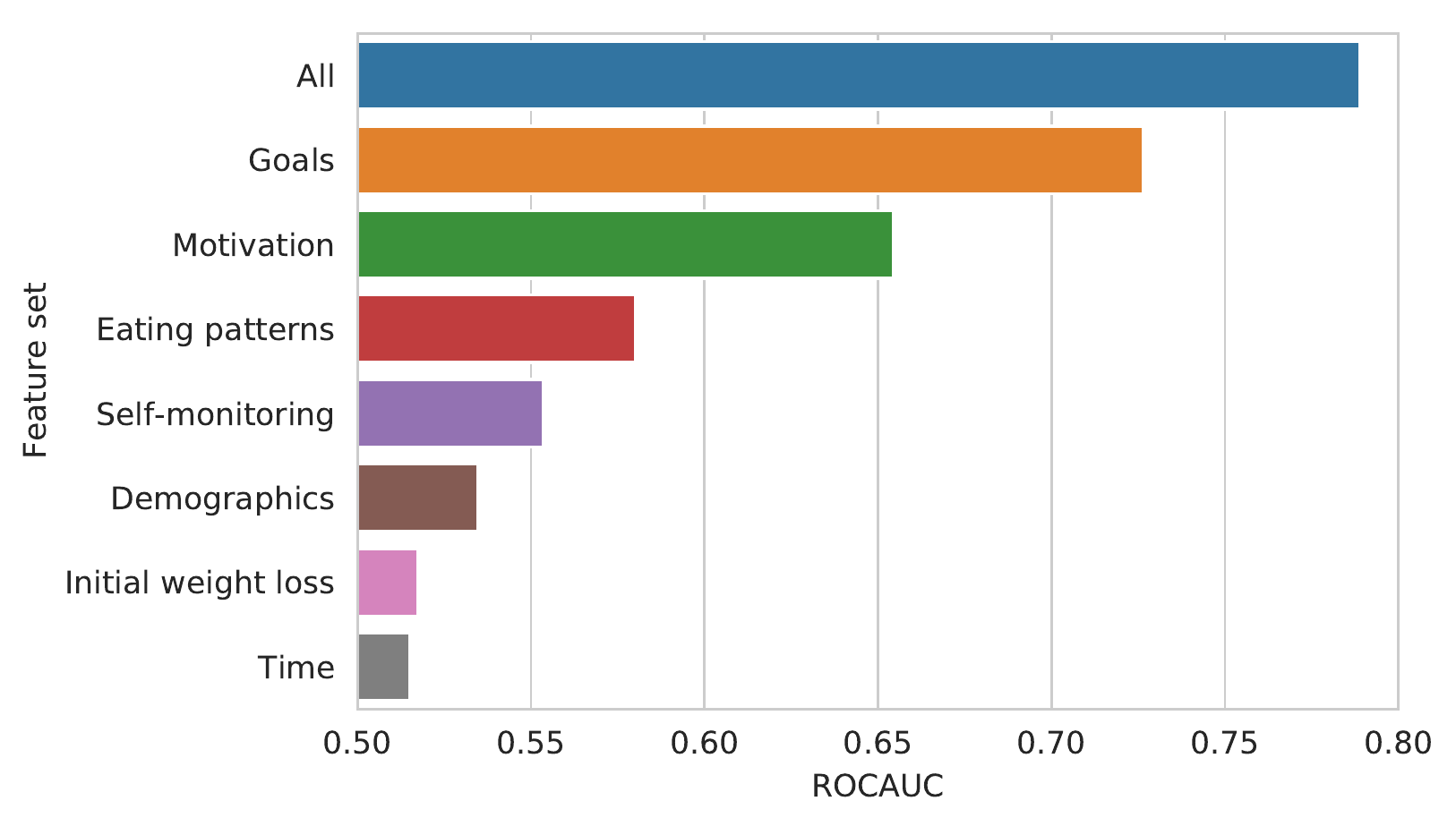}
    \caption{ROC AUC for predicting whether a user will achieve their weight loss goal from observing their behavior over the first seven days after setting it. We report accuracy for each feature set individually and all features together. We reach 79\% ROC AUC in a combined model with all features.}
    \label{fig:predresults}
\end{figure}

Figure \ref{fig:predresults} shows the prediction accuracy of our models. With a model trained on all available features, we achieve a an accuracy of 79\% ROC AUC. 

We find the most predictive features to be the initial goal and the user's motivation. Strikingly, motivation is significantly more predictive than eating patterns. We note that our motivation model was trained with a simple TF-IDF analysis of the words a user writes in their short free-text goal justification. Future work should explore more sophisticated models and more deeply analyze how free-text motivation for a goal can predict goal achievement rates. Surprisingly, we see that initial weight loss on its own was not a strong predictor of propensity to meet a goal, which is likely explained by the fact that initial loss is somewhat meaningless without knowing the goal that the loss is working towards.

\section{Discussion}
Goals are a critical component of activity tracking apps. However, long-term goals such as weight loss are rarely met. This paper moves towards addressing this challenge by analyzing user behavior related to both setting and achieving their goals. In particular, we focus on how behavior during the first week after setting a goal predicts whether or not that goal will be achieved. We do so because early detection and notification of goals that are too hard may enable a new class of interventions where users can be encouraged make their goals easier before becoming frustrated and leaving the app. We identify a set of factors which predict goal achievement related to demographics, initial weight loss, self-monitoring, dietary intake, and motivation, summarized in Section 1.

\xhdr{Limitations}
There are several important limitations to our work, primarily related to generalizability and causality. First, we note that a weight loss goal is a long-term health goal, which we saw typically takes weeks, months, or even years to achieve. It is an open question as to whether the characteristics of users who achieve short-term goals, such as ``walk 10 minutes today'', might differ. 

We also note that our results are correlational in nature. Ideally, the factors we analyze would represent causal effects to help us understand how to improve weight loss goal achievement. While in our analysis we both build off of prior causal findings and make efforts to reduce potential confounders, the factors that lead to weight loss are enormously complex and causal analysis is often extremely difficult or impossible. However, this is not a necessary requirement for such factors to be useful in machine learning models that predict whether or not a goal is likely to be met. We show that even simple models can predict goal achievement after observing just one week of behavior with reasonably high accuracy.

\xhdr{Implications And Future Work}
In addition to uncovering important factors in how people set and achieve their weight loss goals in an activity tracking app, our work also suggests several practical implications. Our work shows that it is possible to predict whether a user will achieve their weight loss goals after observing just their first week of behavior while attempting to meet that goal. This opens the door for activity tracking apps to build a new type of intervention which notifies a user that they seem unlikely to meet their long term goal. Such an intervention might enable users to decide that they should set an easier goal, before they become frustrated and drop out of the app.

Future work should consider several important directions to determine both how to create such an intervention and to determine the effectiveness of such an intervention. First, work should investigate what a activity tracking app do when it recognizes that a goal will never be met, and how users respond to potential interventions. For instance, it is possible than if the user is told too forcefully that they will never meet their goal, perhaps they'll just immediately give up without aiming for an easier goal. On the other hand, if a user is simply assigned an easier goal, it is possible that they will internally still be interested in meeting their first goal, and thus do not truly have an easier one.

\section{Conclusion}
We present a quantitative analysis of the weight loss goals that people set in MyFitnessPal, and the early behaviors that lead to the achievement of those goals. We find that women set more ambitious goals than men, and that younger users set more ambitious goals than older users. We find that most of these goals are never achieved, but that easier goals are far more likely to be achieved than harder ones. We then show that propensity to meet a goal varies significantly by behavior observed during the first week after setting the goal, despite the fact that goals often take months or even years to achieve. We find that, contrary to what findings from medical studies on weight-loss maintenance might suggest, there is no amount of early weight loss which is ``too much''; rather more early progress towards a goal predicts a higher propensity to achieve that goal. We also find that people who log more calories are less likely to meet their goal, but only when we account for commitment to logging. We introduce a novel application of topic modeling to show the four primary motivations that users have for selecting their goals. Finally, we show that the results presented in our paper are sufficiently strong to predict, after observing seven days of behavior, whether a user will ultimately meet their goal, with an accuracy of up to 79\% ROC AUC. Our findings suggest that activity tracking apps may be able to help their users avoid the pitfalls of failing to meet their goals by asking them them select a more realistic goal after observing their holistic behavior and motivations early-on.

\section*{Acknowledgements}
We thank Claire Donnat, Rok Sosic, Emma Pierson, and MyFitnessPal. M.G. was supported by a Junglee Corporation Stanford Graduate Fellowship. This research has been supported in part by 
NSF OAC-1835598,
ARO MURI, 
Stanford Data Science Initiative,
and Chan Zuckerberg Biohub.
\bibliographystyle{ACM-Reference-Format}
\bibliography{refs}


\begin{thebibliography}{50}


\ifx \showCODEN    \undefined \def \showCODEN     #1{\unskip}     \fi
\ifx \showDOI      \undefined \def \showDOI       #1{#1}\fi
\ifx \showISBNx    \undefined \def \showISBNx     #1{\unskip}     \fi
\ifx \showISBNxiii \undefined \def \showISBNxiii  #1{\unskip}     \fi
\ifx \showISSN     \undefined \def \showISSN      #1{\unskip}     \fi
\ifx \showLCCN     \undefined \def \showLCCN      #1{\unskip}     \fi
\ifx \shownote     \undefined \def \shownote      #1{#1}          \fi
\ifx \showarticletitle \undefined \def \showarticletitle #1{#1}   \fi
\ifx \showURL      \undefined \def \showURL       {\relax}        \fi
\providecommand\bibfield[2]{#2}
\providecommand\bibinfo[2]{#2}
\providecommand\natexlab[1]{#1}
\providecommand\showeprint[2][]{arXiv:#2}

\bibitem[\protect\citeauthoryear{Adams, Sallis, Norman, Hovell, Hekler, and
  Perata}{Adams et~al\mbox{.}}{2013}]%
        {adams2013adaptive}
\bibfield{author}{\bibinfo{person}{Marc~A Adams}, \bibinfo{person}{James~F
  Sallis}, \bibinfo{person}{Gregory~J Norman}, \bibinfo{person}{Melbourne~F
  Hovell}, \bibinfo{person}{Eric~B Hekler}, {and} \bibinfo{person}{Elyse
  Perata}.} \bibinfo{year}{2013}\natexlab{}.
\newblock \showarticletitle{An adaptive physical activity intervention for
  overweight adults: a randomized controlled trial}.
\newblock \bibinfo{journal}{\emph{PloS one}} \bibinfo{volume}{8},
  \bibinfo{number}{12} (\bibinfo{year}{2013}), \bibinfo{pages}{e82901}.
\newblock


\bibitem[\protect\citeauthoryear{Althoff, Jindal, and Leskovec}{Althoff
  et~al\mbox{.}}{2017}]%
        {althoff2017online}
\bibfield{author}{\bibinfo{person}{Tim Althoff}, \bibinfo{person}{Pranav
  Jindal}, {and} \bibinfo{person}{Jure Leskovec}.}
  \bibinfo{year}{2017}\natexlab{}.
\newblock \showarticletitle{Online actions with offline impact: How online
  social networks influence online and offline user behavior}. In
  \bibinfo{booktitle}{\emph{Proceedings of the Tenth ACM International
  Conference on Web Search and Data Mining}}. ACM, \bibinfo{pages}{537--546}.
\newblock


\bibitem[\protect\citeauthoryear{Astrup and R{\"o}ssner}{Astrup and
  R{\"o}ssner}{2000}]%
        {astrup2000lessons}
\bibfield{author}{\bibinfo{person}{Arne Astrup} {and} \bibinfo{person}{Stephan
  R{\"o}ssner}.} \bibinfo{year}{2000}\natexlab{}.
\newblock \showarticletitle{Lessons from obesity management programmes: greater
  initial weight loss improves long-term maintenance}.
\newblock \bibinfo{journal}{\emph{obesity reviews}} \bibinfo{volume}{1},
  \bibinfo{number}{1} (\bibinfo{year}{2000}), \bibinfo{pages}{17--19}.
\newblock


\bibitem[\protect\citeauthoryear{Bandura}{Bandura}{1977}]%
        {bandura1977self}
\bibfield{author}{\bibinfo{person}{Albert Bandura}.}
  \bibinfo{year}{1977}\natexlab{}.
\newblock \showarticletitle{Self-efficacy: toward a unifying theory of
  behavioral change.}
\newblock \bibinfo{journal}{\emph{Psychological review}} \bibinfo{volume}{84},
  \bibinfo{number}{2} (\bibinfo{year}{1977}), \bibinfo{pages}{191}.
\newblock


\bibitem[\protect\citeauthoryear{Bodenheimer and Handley}{Bodenheimer and
  Handley}{2009}]%
        {bodenheimer2009goal}
\bibfield{author}{\bibinfo{person}{Thomas Bodenheimer} {and}
  \bibinfo{person}{Margaret~A Handley}.} \bibinfo{year}{2009}\natexlab{}.
\newblock \showarticletitle{Goal-setting for behavior change in primary care:
  an exploration and status report}.
\newblock \bibinfo{journal}{\emph{Patient education and counseling}}
  \bibinfo{volume}{76}, \bibinfo{number}{2} (\bibinfo{year}{2009}),
  \bibinfo{pages}{174--180}.
\newblock


\bibitem[\protect\citeauthoryear{Burke, Wang, and Sevick}{Burke
  et~al\mbox{.}}{2011}]%
        {burke2011self}
\bibfield{author}{\bibinfo{person}{Lora~E Burke}, \bibinfo{person}{Jing Wang},
  {and} \bibinfo{person}{Mary~Ann Sevick}.} \bibinfo{year}{2011}\natexlab{}.
\newblock \showarticletitle{Self-monitoring in weight loss: a systematic review
  of the literature}.
\newblock \bibinfo{journal}{\emph{Journal of the American Dietetic
  Association}} \bibinfo{volume}{111}, \bibinfo{number}{1}
  (\bibinfo{year}{2011}), \bibinfo{pages}{92--102}.
\newblock


\bibitem[\protect\citeauthoryear{Choi, Pak, and Choi}{Choi
  et~al\mbox{.}}{2007}]%
        {choi2007daily}
\bibfield{author}{\bibinfo{person}{Bernard~CK Choi}, \bibinfo{person}{Anita~WP
  Pak}, {and} \bibinfo{person}{Jerome~CL Choi}.}
  \bibinfo{year}{2007}\natexlab{}.
\newblock \showarticletitle{Daily step goal of 10,000 steps: a literature
  review}.
\newblock \bibinfo{journal}{\emph{Clinical \& Investigative Medicine}}
  \bibinfo{volume}{30}, \bibinfo{number}{3} (\bibinfo{year}{2007}),
  \bibinfo{pages}{146--151}.
\newblock


\bibitem[\protect\citeauthoryear{Consolvo, Klasnja, McDonald, and
  Landay}{Consolvo et~al\mbox{.}}{2009}]%
        {consolvo2009goal}
\bibfield{author}{\bibinfo{person}{Sunny Consolvo}, \bibinfo{person}{Predrag
  Klasnja}, \bibinfo{person}{David~W McDonald}, {and} \bibinfo{person}{James~A
  Landay}.} \bibinfo{year}{2009}\natexlab{}.
\newblock \showarticletitle{Goal-setting considerations for persuasive
  technologies that encourage physical activity}. In
  \bibinfo{booktitle}{\emph{Proceedings of the 4th international Conference on
  Persuasive Technology}}. ACM, \bibinfo{pages}{8}.
\newblock


\bibitem[\protect\citeauthoryear{Cordeiro, Epstein, Thomaz, Bales, Jagannathan,
  Abowd, and Fogarty}{Cordeiro et~al\mbox{.}}{2015}]%
        {cordeiro2015barriers}
\bibfield{author}{\bibinfo{person}{Felicia Cordeiro}, \bibinfo{person}{Daniel~A
  Epstein}, \bibinfo{person}{Edison Thomaz}, \bibinfo{person}{Elizabeth Bales},
  \bibinfo{person}{Arvind~K Jagannathan}, \bibinfo{person}{Gregory~D Abowd},
  {and} \bibinfo{person}{James Fogarty}.} \bibinfo{year}{2015}\natexlab{}.
\newblock \showarticletitle{Barriers and negative nudges: Exploring challenges
  in food journaling}. In \bibinfo{booktitle}{\emph{Proceedings of the 33rd
  Annual ACM Conference on Human Factors in Computing Systems}}. ACM,
  \bibinfo{pages}{1159--1162}.
\newblock


\bibitem[\protect\citeauthoryear{Davis, Ellsworth, Payne, Hall, West, and
  Nordhagen}{Davis et~al\mbox{.}}{2016}]%
        {davis2016health}
\bibfield{author}{\bibinfo{person}{Siena~F Davis}, \bibinfo{person}{Marisa~A
  Ellsworth}, \bibinfo{person}{Hannah~E Payne}, \bibinfo{person}{Shelby~M
  Hall}, \bibinfo{person}{Joshua~H West}, {and} \bibinfo{person}{Amber~L
  Nordhagen}.} \bibinfo{year}{2016}\natexlab{}.
\newblock \showarticletitle{Health behavior theory in popular calorie counting
  apps: a content analysis}.
\newblock \bibinfo{journal}{\emph{JMIR mHealth and uHealth}}
  \bibinfo{volume}{4}, \bibinfo{number}{1} (\bibinfo{year}{2016}).
\newblock


\bibitem[\protect\citeauthoryear{Dennison, Morrison, Conway, and
  Yardley}{Dennison et~al\mbox{.}}{2013}]%
        {dennison2013opportunities}
\bibfield{author}{\bibinfo{person}{Laura Dennison}, \bibinfo{person}{Leanne
  Morrison}, \bibinfo{person}{Gemma Conway}, {and} \bibinfo{person}{Lucy
  Yardley}.} \bibinfo{year}{2013}\natexlab{}.
\newblock \showarticletitle{Opportunities and challenges for smartphone
  applications in supporting health behavior change: qualitative study}.
\newblock \bibinfo{journal}{\emph{Journal of medical Internet research}}
  \bibinfo{volume}{15}, \bibinfo{number}{4} (\bibinfo{year}{2013}).
\newblock


\bibitem[\protect\citeauthoryear{Dishman, Vandenberg, Motl, Wilson, and
  DeJoy}{Dishman et~al\mbox{.}}{2009}]%
        {dishman2009dose}
\bibfield{author}{\bibinfo{person}{Rod~K Dishman}, \bibinfo{person}{Robert~J
  Vandenberg}, \bibinfo{person}{Robert~W Motl}, \bibinfo{person}{Mark~G
  Wilson}, {and} \bibinfo{person}{David~M DeJoy}.}
  \bibinfo{year}{2009}\natexlab{}.
\newblock \showarticletitle{Dose relations between goal setting, theory-based
  correlates of goal setting and increases in physical activity during a
  workplace trial}.
\newblock \bibinfo{journal}{\emph{Health education research}}
  \bibinfo{volume}{25}, \bibinfo{number}{4} (\bibinfo{year}{2009}),
  \bibinfo{pages}{620--631}.
\newblock


\bibitem[\protect\citeauthoryear{Donaldson and Normand}{Donaldson and
  Normand}{2009}]%
        {donaldson2009using}
\bibfield{author}{\bibinfo{person}{Jeanne~M Donaldson} {and}
  \bibinfo{person}{Matthew~P Normand}.} \bibinfo{year}{2009}\natexlab{}.
\newblock \showarticletitle{Using goal setting, self-monitoring, and feedback
  to increase calorie expenditure in obese adults}.
\newblock \bibinfo{journal}{\emph{Behavioral Interventions}}
  \bibinfo{volume}{24}, \bibinfo{number}{2} (\bibinfo{year}{2009}),
  \bibinfo{pages}{73--83}.
\newblock


\bibitem[\protect\citeauthoryear{Elfhag and R{\"o}ssner}{Elfhag and
  R{\"o}ssner}{2005}]%
        {elfhag2005succeeds}
\bibfield{author}{\bibinfo{person}{Kristina Elfhag} {and}
  \bibinfo{person}{Stephan R{\"o}ssner}.} \bibinfo{year}{2005}\natexlab{}.
\newblock \showarticletitle{Who succeeds in maintaining weight loss? A
  conceptual review of factors associated with weight loss maintenance and
  weight regain}.
\newblock \bibinfo{journal}{\emph{Obesity reviews}} \bibinfo{volume}{6},
  \bibinfo{number}{1} (\bibinfo{year}{2005}), \bibinfo{pages}{67--85}.
\newblock


\bibitem[\protect\citeauthoryear{Epstein, Caraway, Johnston, Ping, Fogarty, and
  Munson}{Epstein et~al\mbox{.}}{2016}]%
        {epstein2016beyond}
\bibfield{author}{\bibinfo{person}{Daniel~A Epstein}, \bibinfo{person}{Monica
  Caraway}, \bibinfo{person}{Chuck Johnston}, \bibinfo{person}{An Ping},
  \bibinfo{person}{James Fogarty}, {and} \bibinfo{person}{Sean~A Munson}.}
  \bibinfo{year}{2016}\natexlab{}.
\newblock \showarticletitle{Beyond abandonment to next steps: understanding and
  designing for life after personal informatics tool use}. In
  \bibinfo{booktitle}{\emph{Proceedings of the 2016 CHI Conference on Human
  Factors in Computing Systems}}. ACM, \bibinfo{pages}{1109--1113}.
\newblock


\bibitem[\protect\citeauthoryear{Foster, Wadden, Vogt, and Brewer}{Foster
  et~al\mbox{.}}{1997}]%
        {foster1997reasonable}
\bibfield{author}{\bibinfo{person}{Gary~D Foster}, \bibinfo{person}{Thomas~A
  Wadden}, \bibinfo{person}{Renee~A Vogt}, {and} \bibinfo{person}{Gail
  Brewer}.} \bibinfo{year}{1997}\natexlab{}.
\newblock \showarticletitle{What is a reasonable weight loss? Patients'
  expectations and evaluations of obesity treatment outcomes.}
\newblock \bibinfo{journal}{\emph{Journal of consulting and clinical
  psychology}} \bibinfo{volume}{65}, \bibinfo{number}{1}
  (\bibinfo{year}{1997}), \bibinfo{pages}{79}.
\newblock


\bibitem[\protect\citeauthoryear{Ghosh and Guha}{Ghosh and Guha}{2013}]%
        {ghosh2013we}
\bibfield{author}{\bibinfo{person}{Debarchana Ghosh} {and}
  \bibinfo{person}{Rajarshi Guha}.} \bibinfo{year}{2013}\natexlab{}.
\newblock \showarticletitle{What are we ‘tweeting’about obesity? Mapping
  tweets with topic modeling and Geographic Information System}.
\newblock \bibinfo{journal}{\emph{Cartography and geographic information
  science}} \bibinfo{volume}{40}, \bibinfo{number}{2} (\bibinfo{year}{2013}),
  \bibinfo{pages}{90--102}.
\newblock


\bibitem[\protect\citeauthoryear{Gollwitzer}{Gollwitzer}{1999}]%
        {gollwitzer1999implementation}
\bibfield{author}{\bibinfo{person}{Peter~M Gollwitzer}.}
  \bibinfo{year}{1999}\natexlab{}.
\newblock \showarticletitle{Implementation intentions: strong effects of simple
  plans.}
\newblock \bibinfo{journal}{\emph{American psychologist}} \bibinfo{volume}{54},
  \bibinfo{number}{7} (\bibinfo{year}{1999}), \bibinfo{pages}{493}.
\newblock


\bibitem[\protect\citeauthoryear{Gollwitzer and Oettingen}{Gollwitzer and
  Oettingen}{1998}]%
        {gollwitzer1998emergence}
\bibfield{author}{\bibinfo{person}{Peter~M Gollwitzer} {and}
  \bibinfo{person}{Gabriele Oettingen}.} \bibinfo{year}{1998}\natexlab{}.
\newblock \showarticletitle{The emergence and implementation of health goals}.
\newblock \bibinfo{journal}{\emph{Psychology and Health}} \bibinfo{volume}{13},
  \bibinfo{number}{4} (\bibinfo{year}{1998}), \bibinfo{pages}{687--715}.
\newblock


\bibitem[\protect\citeauthoryear{Gowin, Cheney, Gwin, and Franklin~Wann}{Gowin
  et~al\mbox{.}}{2015}]%
        {gowin2015health}
\bibfield{author}{\bibinfo{person}{Mary Gowin}, \bibinfo{person}{Marshall
  Cheney}, \bibinfo{person}{Shannon Gwin}, {and} \bibinfo{person}{Taylor
  Franklin~Wann}.} \bibinfo{year}{2015}\natexlab{}.
\newblock \showarticletitle{Health and fitness app use in college students: a
  qualitative study}.
\newblock \bibinfo{journal}{\emph{American Journal of Health Education}}
  \bibinfo{volume}{46}, \bibinfo{number}{4} (\bibinfo{year}{2015}),
  \bibinfo{pages}{223--230}.
\newblock


\bibitem[\protect\citeauthoryear{Herrmanny, Beckmann, Nachbar, Sauer, Ziegler,
  and Dogang{\"u}n}{Herrmanny et~al\mbox{.}}{2016a}]%
        {herrmanny2016using}
\bibfield{author}{\bibinfo{person}{Katja Herrmanny}, \bibinfo{person}{Nils
  Beckmann}, \bibinfo{person}{Katrin Nachbar}, \bibinfo{person}{Hanno Sauer},
  \bibinfo{person}{J{\"u}rgen Ziegler}, {and} \bibinfo{person}{Ayseg{\"u}l
  Dogang{\"u}n}.} \bibinfo{year}{2016}\natexlab{a}.
\newblock \showarticletitle{Using Psychophysiological Parameters to Support
  Users in Setting Effective Activity Goals}. In
  \bibinfo{booktitle}{\emph{Proceedings of the 2016 CHI Conference Extended
  Abstracts on Human Factors in Computing Systems}}. ACM,
  \bibinfo{pages}{1637--1646}.
\newblock


\bibitem[\protect\citeauthoryear{Herrmanny, Ziegler, and
  Dogang{\"u}n}{Herrmanny et~al\mbox{.}}{2016b}]%
        {herrmanny2016supporting}
\bibfield{author}{\bibinfo{person}{Katja Herrmanny},
  \bibinfo{person}{J{\"u}rgen Ziegler}, {and} \bibinfo{person}{Ayseg{\"u}l
  Dogang{\"u}n}.} \bibinfo{year}{2016}\natexlab{b}.
\newblock \showarticletitle{Supporting users in setting effective goals in
  activity tracking}. In \bibinfo{booktitle}{\emph{International Conference on
  Persuasive Technology}}. Springer, \bibinfo{pages}{15--26}.
\newblock


\bibitem[\protect\citeauthoryear{Hollis, Gullion, Stevens, Brantley, Appel,
  Ard, Champagne, Dalcin, Erlinger, Funk, et~al\mbox{.}}{Hollis
  et~al\mbox{.}}{2008}]%
        {hollis2008weight}
\bibfield{author}{\bibinfo{person}{Jack~F Hollis}, \bibinfo{person}{Christina~M
  Gullion}, \bibinfo{person}{Victor~J Stevens}, \bibinfo{person}{Phillip~J
  Brantley}, \bibinfo{person}{Lawrence~J Appel}, \bibinfo{person}{Jamy~D Ard},
  \bibinfo{person}{Catherine~M Champagne}, \bibinfo{person}{Arlene Dalcin},
  \bibinfo{person}{Thomas~P Erlinger}, \bibinfo{person}{Kristine Funk},
  {et~al\mbox{.}}} \bibinfo{year}{2008}\natexlab{}.
\newblock \showarticletitle{Weight loss during the intensive intervention phase
  of the weight-loss maintenance trial}.
\newblock \bibinfo{journal}{\emph{American journal of preventive medicine}}
  \bibinfo{volume}{35}, \bibinfo{number}{2} (\bibinfo{year}{2008}),
  \bibinfo{pages}{118--126}.
\newblock


\bibitem[\protect\citeauthoryear{Jeffrey, Wing, and Mayer}{Jeffrey
  et~al\mbox{.}}{1998}]%
        {jeffrey1998smaller}
\bibfield{author}{\bibinfo{person}{Robert~W Jeffrey}, \bibinfo{person}{Rena~R
  Wing}, {and} \bibinfo{person}{Randall~R Mayer}.}
  \bibinfo{year}{1998}\natexlab{}.
\newblock \showarticletitle{Are smaller weight losses or more achievable weight
  loss goals better in the long term for obese patients?}
\newblock \bibinfo{journal}{\emph{Journal of consulting and clinical
  psychology}} \bibinfo{volume}{66}, \bibinfo{number}{4}
  (\bibinfo{year}{1998}), \bibinfo{pages}{641}.
\newblock


\bibitem[\protect\citeauthoryear{Laing, Mangione, Tseng, Leng, Vaisberg,
  Mahida, Bholat, Glazier, Morisky, and Bell}{Laing et~al\mbox{.}}{2014}]%
        {laing2014effectiveness}
\bibfield{author}{\bibinfo{person}{Brian~Yoshio Laing},
  \bibinfo{person}{Carol~M Mangione}, \bibinfo{person}{Chi-Hong Tseng},
  \bibinfo{person}{Mei Leng}, \bibinfo{person}{Ekaterina Vaisberg},
  \bibinfo{person}{Megha Mahida}, \bibinfo{person}{Michelle Bholat},
  \bibinfo{person}{Eve Glazier}, \bibinfo{person}{Donald~E Morisky}, {and}
  \bibinfo{person}{Douglas~S Bell}.} \bibinfo{year}{2014}\natexlab{}.
\newblock \showarticletitle{Effectiveness of a smartphone application for
  weight loss compared with usual Care in Overweight Primary Care PatientsA
  randomized, controlled TrialSmartphone application for weight loss in
  overweight primary care patients}.
\newblock \bibinfo{journal}{\emph{Annals of internal medicine}}
  \bibinfo{volume}{161}, \bibinfo{number}{10\_Supplement}
  (\bibinfo{year}{2014}), \bibinfo{pages}{S5--S12}.
\newblock


\bibitem[\protect\citeauthoryear{Linde, Jeffery, Levy, Pronk, and Boyle}{Linde
  et~al\mbox{.}}{2005}]%
        {linde2005weight}
\bibfield{author}{\bibinfo{person}{JA Linde}, \bibinfo{person}{RW Jeffery},
  \bibinfo{person}{RL Levy}, \bibinfo{person}{NP Pronk}, {and}
  \bibinfo{person}{RG Boyle}.} \bibinfo{year}{2005}\natexlab{}.
\newblock \showarticletitle{Weight loss goals and treatment outcomes among
  overweight men and women enrolled in a weight loss trial}.
\newblock \bibinfo{journal}{\emph{International journal of obesity}}
  \bibinfo{volume}{29}, \bibinfo{number}{8} (\bibinfo{year}{2005}),
  \bibinfo{pages}{1002}.
\newblock


\bibitem[\protect\citeauthoryear{Linde, Jeffery, Finch, Ng, and Rothman}{Linde
  et~al\mbox{.}}{2004}]%
        {linde2004unrealistic}
\bibfield{author}{\bibinfo{person}{Jennifer~A Linde}, \bibinfo{person}{Robert~W
  Jeffery}, \bibinfo{person}{Emily~A Finch}, \bibinfo{person}{Debbie~M Ng},
  {and} \bibinfo{person}{Alexander~J Rothman}.}
  \bibinfo{year}{2004}\natexlab{}.
\newblock \showarticletitle{Are unrealistic weight loss goals associated with
  outcomes for overweight women?}
\newblock \bibinfo{journal}{\emph{Obesity Research}} \bibinfo{volume}{12},
  \bibinfo{number}{3} (\bibinfo{year}{2004}), \bibinfo{pages}{569--576}.
\newblock


\bibitem[\protect\citeauthoryear{Locke and Latham}{Locke and Latham}{2002}]%
        {locke2002building}
\bibfield{author}{\bibinfo{person}{Edwin~A Locke} {and} \bibinfo{person}{Gary~P
  Latham}.} \bibinfo{year}{2002}\natexlab{}.
\newblock \showarticletitle{Building a practically useful theory of goal
  setting and task motivation: A 35-year odyssey.}
\newblock \bibinfo{journal}{\emph{American psychologist}} \bibinfo{volume}{57},
  \bibinfo{number}{9} (\bibinfo{year}{2002}), \bibinfo{pages}{705}.
\newblock


\bibitem[\protect\citeauthoryear{Mansi, Milosavljevic, Tumilty, Hendrick,
  Higgs, and Baxter}{Mansi et~al\mbox{.}}{2015}]%
        {mansi2015investigating}
\bibfield{author}{\bibinfo{person}{Suliman Mansi}, \bibinfo{person}{Stephan
  Milosavljevic}, \bibinfo{person}{Steve Tumilty}, \bibinfo{person}{Paul
  Hendrick}, \bibinfo{person}{Chris Higgs}, {and} \bibinfo{person}{David~G
  Baxter}.} \bibinfo{year}{2015}\natexlab{}.
\newblock \showarticletitle{Investigating the effect of a 3-month
  workplace-based pedometer-driven walking programme on health-related quality
  of life in meat processing workers: a feasibility study within a randomized
  controlled trial}.
\newblock \bibinfo{journal}{\emph{BMC public health}} \bibinfo{volume}{15},
  \bibinfo{number}{1} (\bibinfo{year}{2015}), \bibinfo{pages}{410}.
\newblock


\bibitem[\protect\citeauthoryear{McGuire, Wing, Klem, Lang, and Hill}{McGuire
  et~al\mbox{.}}{1999}]%
        {mcguire1999predicts}
\bibfield{author}{\bibinfo{person}{Maureen~T McGuire}, \bibinfo{person}{Rena~R
  Wing}, \bibinfo{person}{Mary~L Klem}, \bibinfo{person}{Wei Lang}, {and}
  \bibinfo{person}{James~O Hill}.} \bibinfo{year}{1999}\natexlab{}.
\newblock \showarticletitle{What predicts weight regain in a group of
  successful weight losers?}
\newblock \bibinfo{journal}{\emph{Journal of consulting and clinical
  psychology}} \bibinfo{volume}{67}, \bibinfo{number}{2}
  (\bibinfo{year}{1999}), \bibinfo{pages}{177}.
\newblock


\bibitem[\protect\citeauthoryear{Mercer, Li, Giangregorio, Burns, and
  Grindrod}{Mercer et~al\mbox{.}}{2016}]%
        {mercer2016behavior}
\bibfield{author}{\bibinfo{person}{Kathryn Mercer}, \bibinfo{person}{Melissa
  Li}, \bibinfo{person}{Lora Giangregorio}, \bibinfo{person}{Catherine Burns},
  {and} \bibinfo{person}{Kelly Grindrod}.} \bibinfo{year}{2016}\natexlab{}.
\newblock \showarticletitle{Behavior change techniques present in wearable
  activity trackers: a critical analysis}.
\newblock \bibinfo{journal}{\emph{JMIR mHealth and uHealth}}
  \bibinfo{volume}{4}, \bibinfo{number}{2} (\bibinfo{year}{2016}).
\newblock


\bibitem[\protect\citeauthoryear{Munson and Consolvo}{Munson and
  Consolvo}{2012}]%
        {munson2012exploring}
\bibfield{author}{\bibinfo{person}{Sean~A Munson} {and} \bibinfo{person}{Sunny
  Consolvo}.} \bibinfo{year}{2012}\natexlab{}.
\newblock \showarticletitle{Exploring goal-setting, rewards, self-monitoring,
  and sharing to motivate physical activity}. In
  \bibinfo{booktitle}{\emph{Pervasive computing technologies for healthcare
  (PervasiveHealth), 2012 6th international conference on}}. IEEE,
  \bibinfo{pages}{25--32}.
\newblock


\bibitem[\protect\citeauthoryear{Nahum-Shani, Smith, Spring, Collins,
  Witkiewitz, Tewari, and Murphy}{Nahum-Shani et~al\mbox{.}}{2017}]%
        {nahum2017just}
\bibfield{author}{\bibinfo{person}{Inbal Nahum-Shani},
  \bibinfo{person}{Shawna~N Smith}, \bibinfo{person}{Bonnie~J Spring},
  \bibinfo{person}{Linda~M Collins}, \bibinfo{person}{Katie Witkiewitz},
  \bibinfo{person}{Ambuj Tewari}, {and} \bibinfo{person}{Susan~A Murphy}.}
  \bibinfo{year}{2017}\natexlab{}.
\newblock \showarticletitle{Just-in-time adaptive interventions (JITAIs) in
  mobile health: key components and design principles for ongoing health
  behavior support}.
\newblock \bibinfo{journal}{\emph{Annals of Behavioral Medicine}}
  \bibinfo{volume}{52}, \bibinfo{number}{6} (\bibinfo{year}{2017}),
  \bibinfo{pages}{446--462}.
\newblock


\bibitem[\protect\citeauthoryear{Normand}{Normand}{2008}]%
        {normand2008increasing}
\bibfield{author}{\bibinfo{person}{Matthew~P Normand}.}
  \bibinfo{year}{2008}\natexlab{}.
\newblock \showarticletitle{Increasing physical activity through
  self-monitoring, goal setting, and feedback}.
\newblock \bibinfo{journal}{\emph{Behavioral Interventions}}
  \bibinfo{volume}{23}, \bibinfo{number}{4} (\bibinfo{year}{2008}),
  \bibinfo{pages}{227--236}.
\newblock


\bibitem[\protect\citeauthoryear{Nothwehr and Yang}{Nothwehr and Yang}{2006}]%
        {nothwehr2006goal}
\bibfield{author}{\bibinfo{person}{Faryle Nothwehr} {and}
  \bibinfo{person}{Jingzhen Yang}.} \bibinfo{year}{2006}\natexlab{}.
\newblock \showarticletitle{Goal setting frequency and the use of behavioral
  strategies related to diet and physical activity}.
\newblock \bibinfo{journal}{\emph{Health education research}}
  \bibinfo{volume}{22}, \bibinfo{number}{4} (\bibinfo{year}{2006}),
  \bibinfo{pages}{532--538}.
\newblock


\bibitem[\protect\citeauthoryear{Olander, Fletcher, Williams, Atkinson, Turner,
  and French}{Olander et~al\mbox{.}}{2013}]%
        {olander2013most}
\bibfield{author}{\bibinfo{person}{Ellinor~K Olander}, \bibinfo{person}{Helen
  Fletcher}, \bibinfo{person}{Stefanie Williams}, \bibinfo{person}{Lou
  Atkinson}, \bibinfo{person}{Andrew Turner}, {and} \bibinfo{person}{David~P
  French}.} \bibinfo{year}{2013}\natexlab{}.
\newblock \showarticletitle{What are the most effective techniques in changing
  obese individuals’ physical activity self-efficacy and behaviour: a
  systematic review and meta-analysis}.
\newblock \bibinfo{journal}{\emph{International Journal of Behavioral Nutrition
  and Physical Activity}} \bibinfo{volume}{10}, \bibinfo{number}{1}
  (\bibinfo{year}{2013}), \bibinfo{pages}{29}.
\newblock


\bibitem[\protect\citeauthoryear{op~den Akker, Jones, and Hermens}{op~den Akker
  et~al\mbox{.}}{2014}]%
        {op2014tailoring}
\bibfield{author}{\bibinfo{person}{Harm op~den Akker},
  \bibinfo{person}{Valerie~M Jones}, {and} \bibinfo{person}{Hermie~J Hermens}.}
  \bibinfo{year}{2014}\natexlab{}.
\newblock \showarticletitle{Tailoring real-time physical activity coaching
  systems: a literature survey and model}.
\newblock \bibinfo{journal}{\emph{User modeling and user-adapted interaction}}
  \bibinfo{volume}{24}, \bibinfo{number}{5} (\bibinfo{year}{2014}),
  \bibinfo{pages}{351--392}.
\newblock


\bibitem[\protect\citeauthoryear{op~den Akker, Klaassen, op~den Akker, Jones,
  and Hermens}{op~den Akker et~al\mbox{.}}{2013}]%
        {op2013opportunities}
\bibfield{author}{\bibinfo{person}{Harm op~den Akker}, \bibinfo{person}{Randy
  Klaassen}, \bibinfo{person}{Rieks op~den Akker}, \bibinfo{person}{Valerie~M
  Jones}, {and} \bibinfo{person}{Hermie~J Hermens}.}
  \bibinfo{year}{2013}\natexlab{}.
\newblock \showarticletitle{Opportunities for smart \& tailored activity
  coaching}. In \bibinfo{booktitle}{\emph{Computer-Based Medical Systems
  (CBMS), 2013 IEEE 26th International Symposium on}}. IEEE,
  \bibinfo{pages}{546--547}.
\newblock


\bibitem[\protect\citeauthoryear{Paul and Dredze}{Paul and Dredze}{2014}]%
        {paul2014discovering}
\bibfield{author}{\bibinfo{person}{Michael~J Paul} {and} \bibinfo{person}{Mark
  Dredze}.} \bibinfo{year}{2014}\natexlab{}.
\newblock \showarticletitle{Discovering health topics in social media using
  topic models}.
\newblock \bibinfo{journal}{\emph{PloS one}} \bibinfo{volume}{9},
  \bibinfo{number}{8} (\bibinfo{year}{2014}), \bibinfo{pages}{e103408}.
\newblock


\bibitem[\protect\citeauthoryear{Pearson}{Pearson}{2012}]%
        {pearson2012goal}
\bibfield{author}{\bibinfo{person}{Erin~S Pearson}.}
  \bibinfo{year}{2012}\natexlab{}.
\newblock \showarticletitle{Goal setting as a health behavior change strategy
  in overweight and obese adults: a systematic literature review examining
  intervention components}.
\newblock \bibinfo{journal}{\emph{Patient education and counseling}}
  \bibinfo{volume}{87}, \bibinfo{number}{1} (\bibinfo{year}{2012}),
  \bibinfo{pages}{32--42}.
\newblock


\bibitem[\protect\citeauthoryear{Pedregosa, Varoquaux, Gramfort, Michel,
  Thirion, Grisel, Blondel, Prettenhofer, Weiss, Dubourg, Vanderplas, Passos,
  Cournapeau, Brucher, Perrot, and Duchesnay}{Pedregosa et~al\mbox{.}}{2011}]%
        {scikit-learn}
\bibfield{author}{\bibinfo{person}{F. Pedregosa}, \bibinfo{person}{G.
  Varoquaux}, \bibinfo{person}{A. Gramfort}, \bibinfo{person}{V. Michel},
  \bibinfo{person}{B. Thirion}, \bibinfo{person}{O. Grisel},
  \bibinfo{person}{M. Blondel}, \bibinfo{person}{P. Prettenhofer},
  \bibinfo{person}{R. Weiss}, \bibinfo{person}{V. Dubourg}, \bibinfo{person}{J.
  Vanderplas}, \bibinfo{person}{A. Passos}, \bibinfo{person}{D. Cournapeau},
  \bibinfo{person}{M. Brucher}, \bibinfo{person}{M. Perrot}, {and}
  \bibinfo{person}{E. Duchesnay}.} \bibinfo{year}{2011}\natexlab{}.
\newblock \showarticletitle{Scikit-learn: Machine Learning in {P}ython}.
\newblock \bibinfo{journal}{\emph{Journal of Machine Learning Research}}
  \bibinfo{volume}{12} (\bibinfo{year}{2011}), \bibinfo{pages}{2825--2830}.
\newblock


\bibitem[\protect\citeauthoryear{Rabbi, Aung, Zhang, and Choudhury}{Rabbi
  et~al\mbox{.}}{2015}]%
        {rabbi2015mybehavior}
\bibfield{author}{\bibinfo{person}{Mashfiqui Rabbi}, \bibinfo{person}{Min~Hane
  Aung}, \bibinfo{person}{Mi Zhang}, {and} \bibinfo{person}{Tanzeem
  Choudhury}.} \bibinfo{year}{2015}\natexlab{}.
\newblock \showarticletitle{MyBehavior: automatic personalized health feedback
  from user behaviors and preferences using smartphones}. In
  \bibinfo{booktitle}{\emph{Proceedings of the 2015 ACM International Joint
  Conference on Pervasive and Ubiquitous Computing}}. ACM,
  \bibinfo{pages}{707--718}.
\newblock


\bibitem[\protect\citeauthoryear{Rooksby, Rost, Morrison, and Chalmers}{Rooksby
  et~al\mbox{.}}{2014}]%
        {rooksby2014personal}
\bibfield{author}{\bibinfo{person}{John Rooksby}, \bibinfo{person}{Mattias
  Rost}, \bibinfo{person}{Alistair Morrison}, {and}
  \bibinfo{person}{Matthew~Chalmers Chalmers}.}
  \bibinfo{year}{2014}\natexlab{}.
\newblock \showarticletitle{Personal tracking as lived informatics}. In
  \bibinfo{booktitle}{\emph{Proceedings of the 32nd annual ACM conference on
  Human factors in computing systems}}. ACM, \bibinfo{pages}{1163--1172}.
\newblock


\bibitem[\protect\citeauthoryear{Schoeppe, Alley, Van~Lippevelde, Bray,
  Williams, Duncan, and Vandelanotte}{Schoeppe et~al\mbox{.}}{2016}]%
        {schoeppe2016efficacy}
\bibfield{author}{\bibinfo{person}{Stephanie Schoeppe},
  \bibinfo{person}{Stephanie Alley}, \bibinfo{person}{Wendy Van~Lippevelde},
  \bibinfo{person}{Nicola~A Bray}, \bibinfo{person}{Susan~L Williams},
  \bibinfo{person}{Mitch~J Duncan}, {and} \bibinfo{person}{Corneel
  Vandelanotte}.} \bibinfo{year}{2016}\natexlab{}.
\newblock \showarticletitle{Efficacy of interventions that use apps to improve
  diet, physical activity and sedentary behaviour: a systematic review}.
\newblock \bibinfo{journal}{\emph{International Journal of Behavioral Nutrition
  and Physical Activity}} \bibinfo{volume}{13}, \bibinfo{number}{1}
  (\bibinfo{year}{2016}), \bibinfo{pages}{127}.
\newblock


\bibitem[\protect\citeauthoryear{Shilts, Horowitz, and Townsend}{Shilts
  et~al\mbox{.}}{2004}]%
        {shilts2004goal}
\bibfield{author}{\bibinfo{person}{Mical~Kay Shilts}, \bibinfo{person}{Marcel
  Horowitz}, {and} \bibinfo{person}{Marilyn~S Townsend}.}
  \bibinfo{year}{2004}\natexlab{}.
\newblock \showarticletitle{Goal setting as a strategy for dietary and physical
  activity behavior change: a review of the literature}.
\newblock \bibinfo{journal}{\emph{American Journal of Health Promotion}}
  \bibinfo{volume}{19}, \bibinfo{number}{2} (\bibinfo{year}{2004}),
  \bibinfo{pages}{81--93}.
\newblock


\bibitem[\protect\citeauthoryear{Strecher, Seijts, Kok, Latham, Glasgow,
  DeVellis, Meertens, and Bulger}{Strecher et~al\mbox{.}}{1995}]%
        {strecher1995goal}
\bibfield{author}{\bibinfo{person}{Victor~J Strecher},
  \bibinfo{person}{Gerard~H Seijts}, \bibinfo{person}{Gerjo~J Kok},
  \bibinfo{person}{Gary~P Latham}, \bibinfo{person}{Russell Glasgow},
  \bibinfo{person}{Brenda DeVellis}, \bibinfo{person}{Ree~M Meertens}, {and}
  \bibinfo{person}{David~W Bulger}.} \bibinfo{year}{1995}\natexlab{}.
\newblock \showarticletitle{Goal setting as a strategy for health behavior
  change}.
\newblock \bibinfo{journal}{\emph{Health education quarterly}}
  \bibinfo{volume}{22}, \bibinfo{number}{2} (\bibinfo{year}{1995}),
  \bibinfo{pages}{190--200}.
\newblock


\bibitem[\protect\citeauthoryear{Veli{\v{c}}kovi{\'c}, Karazija, Lane,
  Bhattacharya, Liberis, Li{\`o}, Chieh, Bellahsen, and
  Vegreville}{Veli{\v{c}}kovi{\'c} et~al\mbox{.}}{2018}]%
        {velivckovic2018cross}
\bibfield{author}{\bibinfo{person}{Petar Veli{\v{c}}kovi{\'c}},
  \bibinfo{person}{Laurynas Karazija}, \bibinfo{person}{Nicholas~D Lane},
  \bibinfo{person}{Sourav Bhattacharya}, \bibinfo{person}{Edgar Liberis},
  \bibinfo{person}{Pietro Li{\`o}}, \bibinfo{person}{Angela Chieh},
  \bibinfo{person}{Otmane Bellahsen}, {and} \bibinfo{person}{Matthieu
  Vegreville}.} \bibinfo{year}{2018}\natexlab{}.
\newblock \showarticletitle{Cross-modal recurrent models for weight objective
  prediction from multimodal time-series data}. In
  \bibinfo{booktitle}{\emph{Proceedings of the 12th EAI International
  Conference on Pervasive Computing Technologies for Healthcare}}. ACM,
  \bibinfo{pages}{178--186}.
\newblock


\bibitem[\protect\citeauthoryear{Wen, Zhang, Liu, and Lei}{Wen
  et~al\mbox{.}}{2017}]%
        {wen2017evaluating}
\bibfield{author}{\bibinfo{person}{Dong Wen}, \bibinfo{person}{Xingting Zhang},
  \bibinfo{person}{Xingyu Liu}, {and} \bibinfo{person}{Jianbo Lei}.}
  \bibinfo{year}{2017}\natexlab{}.
\newblock \showarticletitle{Evaluating the Consistency of Current Mainstream
  Wearable Devices in Health Monitoring: A Comparison Under Free-Living
  Conditions}.
\newblock \bibinfo{journal}{\emph{Journal of medical Internet research}}
  \bibinfo{volume}{19}, \bibinfo{number}{3} (\bibinfo{year}{2017}).
\newblock


\bibitem[\protect\citeauthoryear{Williamson, Serdula, Anda, Levy, and
  Byers}{Williamson et~al\mbox{.}}{1992}]%
        {williamson1992weight}
\bibfield{author}{\bibinfo{person}{David~F Williamson}, \bibinfo{person}{Mary~K
  Serdula}, \bibinfo{person}{Robert~F Anda}, \bibinfo{person}{Alan Levy}, {and}
  \bibinfo{person}{Tim Byers}.} \bibinfo{year}{1992}\natexlab{}.
\newblock \showarticletitle{Weight loss attempts in adults: goals, duration,
  and rate of weight loss.}
\newblock \bibinfo{journal}{\emph{American Journal of Public Health}}
  \bibinfo{volume}{82}, \bibinfo{number}{9} (\bibinfo{year}{1992}),
  \bibinfo{pages}{1251--1257}.
\newblock


\bibitem[\protect\citeauthoryear{Wing and Phelan}{Wing and Phelan}{2005}]%
        {wing2005long}
\bibfield{author}{\bibinfo{person}{Rena~R Wing} {and} \bibinfo{person}{Suzanne
  Phelan}.} \bibinfo{year}{2005}\natexlab{}.
\newblock \showarticletitle{Long-term weight loss maintenance--}.
\newblock \bibinfo{journal}{\emph{The American journal of clinical nutrition}}
  \bibinfo{volume}{82}, \bibinfo{number}{1} (\bibinfo{year}{2005}),
  \bibinfo{pages}{222S--225S}.
\newblock


\end{thebibliography}

\end{document}